\documentclass[12pt]{article}
\pdfoutput=1

\usepackage[reqno]{amsmath}
\usepackage{mathrsfs}
\usepackage{amssymb}
\usepackage{multirow}

\usepackage{subfigure}

\usepackage{feynmp}
\usepackage{feynmp-auto}
\usepackage{bbm}
\usepackage{epsfig}
\usepackage{array}
\usepackage{float}
\usepackage{color}
\usepackage{graphicx}

\usepackage{a4}
\usepackage{a4wide}
\usepackage{wasysym}
\usepackage{url}
\usepackage{color}

\usepackage[colorlinks=true,linkcolor=red,citecolor=green,filecolor=magenta,urlcolor=cyan,bookmarks=true,bookmarksopen=true]{hyperref}
\usepackage[all]{hypcap}
\usepackage{cite}

\begin{document}

\begin{titlepage}
\title{\vspace*{-2.0cm}
\hfill {\normalsize MPI-2016-156}\\[3mm]
\bf\Large A Consistent Theory of Decaying Dark Matter Connecting IceCube to the Sesame Street
\\[1mm] }

\author{
Marco Chianese$^{a,b,c}$\thanks{email: \tt chianese@na.infn.it}~~~and~~~~Alexander Merle$^c$\thanks{email: \tt amerle@mpp.mpg.de}
\\ \\
$^a${\normalsize \it INFN, Sezione di Napoli,}\\
{\normalsize \it Complesso Univ. Monte S. Angelo, I-80126 Napoli, Italy}\\
$^b${\normalsize \it Dipartimento di Fisica {\it Ettore Pancini}, Universit\`a di Napoli Federico II,}\\
{\normalsize \it Complesso Univ. Monte S. Angelo, I-80126 Napoli, Italy}\\
$^c${\normalsize \it Max-Planck-Institut f\"ur Physik (Werner-Heisenberg-Institut),}\\
{\normalsize \it F\"ohringer Ring 6, 80805 M\"unchen, Germany}\\
}
\date{\today}
\maketitle
\thispagestyle{empty}

\begin{abstract}
\noindent
The high energy events observed at the IceCube Neutrino Observatory have triggered many investigations interpreting the highly energetic neutrinos detected as decay products of heavy unstable Dark Matter particles. However, while very detailed treatments of the IceCube phenomenology exist, only a few references focus on the (non-trivial) Dark Matter production part -- and all of those rely on relatively complicated new models which are not always testable directly. We instead investigate two of the most minimal scenarios possible, where the operator responsible for the IceCube events is directly involved in Dark Matter production. We show that the simplest (four-dimensional) operator is not powerful enough to accommodate all constraints. A more non-minimal setting (at mass dimension six), however, can do both fitting all the data and also allowing for a comparatively small parameter space only, parts of which can be in reach of future observations. We conclude that minimalistic approaches can be enough to explain all data required, while complicated new physics seems not to be required by IceCube.
\end{abstract}

\end{titlepage}

\section{\label{sec:intro}Introduction}

The IceCube Neutrino Observatory, a neutrino telescope located at the Amundsen-Scott South Pole Station, is a unique window to observe highly energetic neutrinos reaching the Earth's surface, originating from sources as close as the upper regions of the atmosphere up to extra-galactic objects~\cite{Waxman:2014vja}. Its applications to closer sources range from a more precise determination of the atmospheric neutrino flux~\cite{Aartsen:2015rwa} over measuring the properties of active neutrinos~\cite{Aartsen:2014yll} and constraining those of sterile neutrinos~\cite{TheIceCube:2016oqi} to astrophysical findings such as the shadowing effect of the moon on cosmic rays~\cite{MoonShadow}. As for the wider sources, IceCube's goal is to investigate several types of astrophysical neutrino emitters, its possible applications ranging from astrophysical point sources~\cite{Aartsen:2016tpb} over Dark Matter annihilation~\cite{Aartsen:2016pfc} to supernovae~\cite{Aartsen:2015trq}. Finally, also certain exotic particles may leave visible signatures in the detector, such as magnetic monopoles~\cite{Aartsen:2015exf}.

A big surprise in the data taken between 2010 and 2013 was the detections of three very high energy events, reported in Refs.~\cite{Aartsen:2013bka,Aartsen:2013jdh,Aartsen:2014gkd}. These events have been under such scrutiny and have generated such an amount of interest, that they have even been given names after characters of the Sesame Street~\cite{SesameStreet} for better recognition: Ernie (1.14~PeV), Bert (1.04~PeV), and Big Bird (2.2~PeV).

The origin of these very high energy events is still unclear, though. The initial discussion was immediately targeting various astrophysical sources, see Refs.~\cite{Cholis:2012kq,Anchordoqui:2013dnh,Murase:2014tsa} for comprehensive treatments and extensive lists of references. However, in the particle physics community, great interest arose instead in relating the detections to the physics of Dark Matter (DM), in order to address one of the most fascinating topics in all of science. It had been argued that such high energy events probably cannot originate from DM \emph{annihilation}~\cite{Bai:2013nga} because of the unitarity bound~\cite{Griest:1989wd,Hui:2001wy}. Thus, although this bound may be circumvented~\cite{Zavala:2014dla}, most works have focused on DM \emph{decay} instead. Looking at the literature, most authors consider the decay of superheavy DM-type particles~\cite{Feldstein:2013kka,Bai:2013nga,Esmaili:2013gha,Ema:2013nda,Bhattacharya:2014vwa,Fong:2014bsa,Rott:2014kfa,Dudas:2014bca,Murase:2015gea,Anchordoqui:2015lqa,Ko:2015nma,Aisati:2015vma,Higaki:2014dwa,Roland:2015yoa,Esmaili:2014rma,Boucenna:2015tra}, although some work has also been presented on lower-mass candidates boosted to high energies~\cite{Bhattacharya:2014yha,Kopp:2015bfa}. In general, depending on the interaction between DM and Standard Model particles, the decays of superheavy DM particles may be able to account for the whole TeV--PeV IceCube diffuse neutrino flux (see for instance Ref.~\cite{Esmaili:2014rma}) -- or at least for part of it, as shown in Ref.~\cite{Boucenna:2015tra}, where the TeV neutrinos events are explained in terms of an astrophysical power-law flux (two-component flux). While all kinds of phenomenological aspects of the signal are considered, like e.g.\ its variation with the DM profile~\cite{Chianese:2016opp}, most settings are not specified very accurately from the particle physics side, making it tempting to unify the treatments based on a set of effective operators mediating DM decay~\cite{Esmaili:2014rma,Boucenna:2015tra}.

Although the IceCube part has been treated in great detail, the literature on how to produce such a type of DM in the first place appears a bit scarce in comparison. Nevertheless, there are some notable exceptions which treat the full course of DM production down to an analysis of the IceCube signal: in the examples found, the DM particles are e.g.\ produced in a secluded sector~\cite{Dev:2016qbd}, by freeze-out with resonantly enhanced annihilations~\cite{DiBari:2016guw}, or via \emph{freeze-in}~\cite{Fiorentin:2016avj,Roland:2015yoa,Fong:2014bsa}.

It is this latter mechanism we would also like to focus on in our current work. While Ref.~\cite{Fiorentin:2016avj} investigated a full model based on left-right symmetry, we go the opposite way and try to be very minimalistic by asking the question which of the possible operators mediating DM decay~\cite{Esmaili:2014rma,Boucenna:2015tra} could \emph{at the same time} be responsible for DM production in the early Universe. We will in particular focus on the 4-dimensional operator discussed in Ref.~\cite{Esmaili:2014rma}, which allows the DM particle to directly decay into a neutrino and a SM Higgs, as well as on an alternative leptophilic 6-dimensional operator which has a somewhat richer phenomenology and features the same predictions as the one discussed in Ref.~\cite{Boucenna:2015tra}. As we will see, while the minimal ($d=4$)-operator is in fact not sufficient to bring DM production in accordance with the IceCube signal (unless both parts are completely disentangled, as in Ref.~\cite{Fiorentin:2016avj}), the ($d=6$)-operator turns out to be powerful enough: not only can it accommodate for all data and bounds, but it actually leaves us with a potentially testable allowed window. We therefore show that, beyond the ingredients needed for DM production and (of course) a candidate DM particle, no complicated new physics is needed to ensure both consistency and testability.

This paper is structured as follows. We start by introducing the basic underlying setup in Sec.~\ref{sec:basic}, before giving a general discussion on the necessary characteristics of decaying DM in Sec.~\ref{sec:general}. DM production with the different operators is discussed in detail in Sec.~\ref{sec:DM-production}, before our numerical results are presented and discussed in Sec.~\ref{sec:Results}. We finally conclude in Sec.~\ref{sec:conc}. Technical details are given in App.~\ref{app:A}, which lists the explicit expressions for all matrix elements used in the computation of DM production.

\section{\label{sec:basic}The basic idea}

Our basic idea is to explicitly compute DM production for two operators that have been used to explain the IceCube high energy signals, namely $\overline{L_L} H \chi$~\cite{Esmaili:2014rma} and $(\overline{L_L}\ell_R)(\overline{L_L}\chi)$~\cite{Boucenna:2015tra}, which both feature a DM particle $\chi$ transforming as $\chi \sim (\mathbf{1}, \mathbf{1}, 0)$ under the Standard Model (SM) gauge group $SU(3)_C \times SU(2)_L \times U(1)_Y$. Note that $\chi$ is basically a right-handed neutrino; however, we would like to keep the discussion general as there may also be settings in which $\chi$ has some further non-trivial charges, and thus there may also be alternative interpretations of $\chi$.

Explicitly, the two different operators are:
\begin{enumerate}

\item The {\bf 4-dimensional operator}~\cite{Esmaili:2014rma}:
\begin{equation}
 y_{\alpha\chi} \overline{L_{L\alpha}} H \chi\,.
 \label{eq:LH}
\end{equation}
Here, $\chi$ is the DM particle introduced above, $H \sim (\mathbf{1}, \mathbf{2}, +1/2)$ is the SM Higgs doublet, and $L_{L\alpha} \sim (\mathbf{1}, \mathbf{2}, -1/2)$ is the left-handed lepton doublet of generation $\alpha$, with $\alpha = e, \mu, \tau$. Note that the operator in Eq.~\eqref{eq:LH} is allowed as soon as all the necessary particles exist.

\item The {\bf 6-dimensional operator} (phenomenologically identical to yet different from the one in Ref.~\cite{Boucenna:2015tra}):
\begin{equation}
 \frac{\lambda_{\alpha\beta} \lambda'_\gamma}{M_S^2} \left(\overline{(L_{L\alpha})^C} i\sigma_2 L_{L\beta}\right) \left(\overline{\ell_{R\gamma}} \chi \right)\,,
 \label{eq:sq-1}
\end{equation}
where the superscript $C$ indicates fermionic charge-conjugation, $\Psi^C =\mathcal{\tilde{C}}\ \overline{\Psi}^T$ with $\mathcal{\tilde{C}} = i \gamma^2 \gamma^0$, $i\sigma_2 = \begin{pmatrix} 0 & 1\\ -1 & 0 \end{pmatrix}$ is a matrix in the $SU(2)_L$ group space, and $\ell_{R\gamma}$ is the right-handed charged lepton field of generation $\gamma$.

In order to compute DM production accurately, we need to find a viable ultraviolet completion behind the effective operator given in Eq.~\eqref{eq:sq-2}. We can do this by introducing an electrically charged but $SU(2)_L$ singlet scalar $S^+ \sim (\mathbf{1}, \mathbf{1}, 1)$ (and its antiparticle). This new scalar features a (potentially) lepton number violating coupling just like that used in the Zee-Babu model~\cite{Zee:1985rj,Zee:1985id,Babu:1988ki}:\footnote{The Zee-Babu model is a model explaining the smallness of neutrino masses, by generating them only at 2-loop level while they are forbidden at lower orders. This model features two $SU(2)$ singlet scalars, one of which is doubly charged while the other one carries a single electric charge, the latter carrying the same quantum numbers as $S^+$ in Eq.~\eqref{eq:sq-2}. This model is particularly interesting in what concerns its lepton flavour violation~\cite{AristizabalSierra:2006gb} and collider phenomenology~\cite{Schmidt:2014zoa,Geib:2015tvt}, which is linked to the light neutrino masses~\cite{Long:2014fja}. Note that, contrary to the setup we use here, the Zee-Babu model does not usually feature total singlet fermion fields, although it can be extended to do so and would then also be able to accommodate for Dark Matter~\cite{Schmidt:2012yg}.}
\begin{equation}
 \lambda_{\alpha\beta} \overline{(L_{L\alpha})^C} i\sigma_2 L_{L\beta} S^+ + h.c.
 \label{eq:sq-2}
\end{equation}
Furthermore, the new particle $S^\pm$ can couple to the DM particle $\chi$ according to
\begin{equation}
 \lambda'_\gamma \overline{\ell_{R\gamma}} \chi S^-  + h.c.
 \label{eq:sq-3}
\end{equation}
Thus, for a very heavy particle $S^\pm$ of mass $M_S$, the effective operator in Eq.~\eqref{eq:sq-1} is generated if the charged scalar is integrated out.
\end{enumerate}

We are thus in a situation where, depending on the values of the couplings $y_{\alpha\chi}$, $\lambda_{\alpha\beta}$, and $\lambda'_\gamma$, as well as on the mass $M_S$, it could very well be that either the operator in Eq.~\eqref{eq:LH} or the one in Eq.~\eqref{eq:sq-2} can be dominant.\\

For example, we could also have obtained an operator similar to that in Eq.~\eqref{eq:sq-1}, namely $(\overline{L_L}\ell_R)(\overline{L_L} \chi)$, had we simply integrated out the SM-Higgs in case it coupled as in Eq.~\eqref{eq:LH} (this was in fact the operator originally discussed in Ref.~\cite{Boucenna:2015tra}). However, in that case the two operators in Eqs.~\eqref{eq:LH} and~\eqref{eq:sq-1} would not be independent -- instead, the 6-dimensional operator would be induced by the 4-dimensional one and both would contain the coupling $y_{\alpha\chi}$. In particular, the $(d=6)$-operator would feature a second small coupling and would be subdominant compared to the 4-dimensional one.

On the other hand, we can also generate a situation in which the $(d=6)$-operator can dominate over the $(d=4)$-operator, and which contains the simple setting discussed. Let us assume an $A_4$ symmetry (see, e.g., Refs.~\cite{Altarelli:2010gt,Ishimori:2010au,King:2013eh,King:2014nza} for details), and let us take the  following assignment:
\begin{equation}
\begin{array}{c}
L_L= (L_{Le}, L_{L\mu}, L_{L\tau}) \sim \mathbf{3}\,,\quad  \ell_R = (e_R, \mu_R, \tau_R) \sim \mathbf{3}\,, \\\\
H_1 \sim \mathbf{1}\,, \quad H_2 \sim \mathbf{1'} \,, \quad H_3 \sim \mathbf{1''}\,, \\\\
 S^+ = (S^+_1, S^+_2, S^+_3) \sim \mathbf{3} \,,\quad  \chi \sim \mathbf{1} \,,
\end{array}
\label{eq:A4-assignments}
\end{equation}
where we have split both $H$ and $S$ into several components, for the sake of a suitable assignment, which however will not change much (in particular if we set the component masses equal). The three Higgs doublets $H_{1,2,3}$ are required to give different masses to $e$, $\mu$, and $\tau$. Under these assignments, the operator $\overline{L_L} H_{(1,2,3)} \chi$ is forbidden at tree-level, because it would transform as $\mathbf{3} \otimes (\mathbf{1}, \mathbf{1'}, \mathbf{1''}) \otimes \mathbf{1} \not\supset \mathbf{1}$, while $ \left(\overline{(L_L)^C} i\sigma_2 L_L\right) \left(\overline{\ell_R} \chi \right) \sim (\mathbf{3} \otimes \mathbf{3}) \otimes (\mathbf{3} \otimes \mathbf{1}) \supset \mathbf{3} \otimes \mathbf{3} \supset \mathbf{1}$. However, once the vacuum expectation values $\langle H_{2,3} \rangle$ break the $A_4$-symmetry, one can generate the operator $\overline{L_L} H_{(1,2,3)} \chi$ at 1-loop level by glueing together the vertices $\overline{\ell_R} \chi S$, $\overline{L_L} H \ell_R$, and $\overline{(L_L)^C} L_L S$. Then, the resulting $(d=4)$-operator only arises at one-loop level and is suppressed by being proportional to the cube of the tiny coupling, whereas the $(d=6)$-operator is only suppressed by its square. Thus, indeed, depending on the situation, one or the other operator might be dominant, and it thus makes sense to discuss both cases in some detail.\\

As a final subtlety we already hinted on, note that the operator in Eq.~\eqref{eq:sq-1} in fact slightly differs from the one used in Ref.~\cite{Boucenna:2015tra}, which would rather be of the form $(\overline{L_L} \ell_R) (\overline{L_L} \chi)$. The reason for this is that Ref.~\cite{Boucenna:2015tra} relied on the earlier classifications of operators presented in Refs.~\cite{Haba:2010ag,delAguila:2008ir}. However, these older references only treated operators which lepton number violation solely originated from the right-handed neutrino sector, while in Eq.~\eqref{eq:sq-2} it has its origin in a new scalar field which prevents the full Lagrangian from being assigned a lepton number. However, in what concerns the IceCube phenomenology, the two operators $(\overline{L_L} \ell_R) (\overline{L_L} \chi)$ and $(\overline{(L_L)^C} i\sigma_2 L_L) (\overline{\ell_R} \chi)$ are, in fact, indistinguishable with the present IceCube accuracy~\cite{Boucenna:2015tra}. Instead, we focus on whether or not the constraints derived from IceCube can be met by the DM produced in the early Universe with any of the two operators.

\section{\label{sec:general}General thoughts on decaying DM and IceCube}

The next point to discuss is the type of DM we would like to investigate, which is restricted by both its production in the early Universe and the IceCube data. The production mechanism we would like to use is freeze-in production~\cite{Hall:2009bx}, see Sec.~\ref{sec:DM-production} for details, in which the DM particles are never in thermal equilibrium but are still feebly coupled to the SM and thus gradually produced from the thermal bath in the early Universe. We should note that, during the final phase of this work, Ref.~\cite{Fiorentin:2016avj} appeared which was thus the first to discuss freeze-in production in connection to the IceCube high energy events. This reference features a full Left-Right symmetric model, in which DM production and the IceCube events arise from different parts of the theory. This is indeed one way to get things consistent. We do however pursue a different path and investigate whether DM production and the IceCube events could arise from \emph{one and the same} interaction. Note further that Ref.~\cite{Fiorentin:2016avj} relied on some couplings being very tiny, which may seem like an unnatural fine tuning at first sight. However, as we will show, a large degree of fine tuning is in fact \emph{unavoidable} for decaying DM of the type accessible at IceCube, which is simply reflected by the settings of both Ref.~\cite{Fiorentin:2016avj} and our work.

Decaying DM is in general somewhat unnatural, in the sense that the lifetime of the DM particles has to be at least larger than the age of the Universe~\cite{Audren:2014bca}, which means that the DM decay necessarily has to be a suppressed process. Now the question arises how we can possibly obtain such strong suppressions, keeping in mind that the decay rate of any particle roughly scales with some power of its mass, unless some conservation law keeps it stable (which is intrinsically not possible for decaying DM).

Various reasons for a suppressed decay rate could be thought of:
\begin{enumerate}
\item {\bf Small phase space:} This can be achieved by either choosing the mass of the decaying particle to be small or to only allow for final state particles whose sum of masses is nearly identical to the mass of the parent particle. The former is employed, for example, for keV sterile neutrinos~\cite{Abazajian:2001vt,Adhikari:2016bei}, while the latter option was e.g.\ used to explain the $3.5$~keV hint~\cite{Bulbul:2014sua,Boyarsky:2014jta} by decays of excited DM states~\cite{Finkbeiner:2014sja,Frandsen:2014lfa}.\\
$\Rightarrow$ \emph{Both these options are not applicable if we want to have very highly energetic final states, as needed to explain the IceCube data.}

\item {\bf Planck-scale suppressed operators:} In some cases, processes that are otherwise forbidden may only be induced at very high energies~\cite{Lew:1993yt}, where gravity is expected to break global symmetries~\cite{Rai:1992xw}. The resulting interaction (or in this case decay) rates are then usually very small.\\
$\Rightarrow$ \emph{This could actually work in the case at hand. However, unless a full UV-complete theory is specified, an introduction of Planck-scale suppressed operators is not much more than a parametrisation of the apparent lack of knowledge.}

\item {\bf Couplings tuned to tiny values:} After all, this is the remaining possibility once other ideas are exhausted. However, given that the previous two possibilities either do not work or are just pushing the problem to different scales, tuning seems to be the final generic option. Or, turning round the logic, any setting explaining the IceCube data via decaying DM will \emph{necessarily} be tuned, unless unknown exotic high-scale physics is assumed, which may alleviate the tension.\\
$\Rightarrow$ \emph{To explain IceCube in terms of decaying DM, tuning actually seems to be the most ``natural'' option: if the high-energy signals are to be explained by DM decay, there will be hardly any way around fine-tuning certain couplings.}
\end{enumerate}
There are in fact no other simple ways to suppress the decay rate, because apart from the initial state mass, from the phase space, and from the size of the squared matrix element, there are simply no other ingredients that could possibly be varied.\\

The next point is to summarise the constraints arising from the requirement of the correct IceCube phenomenology. As for the mass, given that the maximum energy of the IceCube events has been measured to be about $2$~PeV~\cite{Aartsen:2013jdh,Aartsen:2014gkd}, the mass of the DM particle will be constrained to be:
\begin{equation}
 m_\chi \sim 4~{\rm PeV} = 4\cdot 10^6~{\rm GeV} \ \ \ [\sim 5~{\rm PeV} = 5\cdot 10^6~{\rm GeV}],
 \label{eq:DM-mass-constraint}
\end{equation}
as suggested by a 2-body [3-body]\footnote{The estimated mass value of $5$~PeV instead of the naive expectation of $6$ arises from the shape of the 3-body decay spectrum.} decay into two [three] practically massless final states.

Nore that we \emph{assume} the DM lifetime to be $10^{28}$~sec, which is a good benchmark value for the two operators under consideration, but our qualitative results do not depend strongly on this assumption. Using this, the following rough conditions have to be met:
\begin{itemize}
\item for $y_{\alpha\chi} \overline{L_{L\alpha}} H \chi$ \& $m_\chi = 4~{\rm PeV}$, we obtain
\begin{equation}
 |y_{\alpha\chi}| \sim 1.8 \times 10^{-29}\,,
 \label{eq:Constraint_d=4}
\end{equation}
in accordance with Eq.~(44) from Ref.~\cite{Higaki:2014dwa}.\footnote{It is worth noticing that there is a typo in the Eq.~(4.2) in Ref.~\cite{Esmaili:2014rma} concerning the size of the coupling. However it does not affect the results of their analysis, since the neutrino flux is inversely proportional to the DM lifetime that has been considered equal to $\mathcal{O}(10^{28})$~sec.}
\item for $\frac{\lambda_{\alpha\beta} \lambda'_\gamma}{M_S^2} \left(\overline{(L_{L\alpha})^C} i\sigma_2 L_{L\beta}\right) \left(\overline{\ell_{R\gamma}} \chi \right)$ \& $m_\chi = 5~{\rm PeV}$, we roughly have
\begin{equation}
 \sqrt{|\lambda_{\alpha\beta} \lambda'_\gamma|} \sim 1.6\cdot 10^{-21}\times \left(\frac{M_S}{\text{1 GeV}}\right) \,,
 \label{eq:Constraint_d=6}
\end{equation}
according to the DM decay width, whose general expression will be given later.
\end{itemize}
These constraints already anticipate the key point which will show up in our analysis: while for the $(d=6)$-operator from Eq.~\eqref{eq:sq-2}, one can adjust two couplings ($\lambda_{\alpha\beta}$ and $\lambda'_\gamma$) and one mass ($M_S$) to meet the IceCube and DM production constraints at the same time, the $(d=4)$-operator from Eq.~\eqref{eq:sq-1} only features one single coupling $y_{\alpha\chi}$ to play with. It can thus be expected that successful DM production should be much harder to achieve in case only this single operator is used. The authors of Ref.~\cite{Fiorentin:2016avj} have recognised this fact, however, in their case no problem arose because of extended gauge interactions being present \emph{in addition} to the $(d=4)$-operator from Eq.~\eqref{eq:sq-1}. We will instead try to stick to the most minimal case possible so that -- apart from the DM mass $m_\chi$ -- we only use the minimal set of new quantities available, i.e., $y_{\alpha\chi}$ for the $(d=4)$-operator and $(\lambda_{\alpha\beta}, \lambda'_\gamma, M_S)$ for the $(d=6)$-operator.

\section{\label{sec:DM-production}Freeze-in production of Dark Matter}

Clearly, if we aim to explain the high energy events at IceCube by DM decay, it is not sufficient to just assume some heavy particle which happens to have the correct abundance and lifetime, but it has to be produced in a suitable way in the early Universe. The most generic production mechanism for WIMP-like DM is the so-called thermal freeze-out~\cite{Lee:1977ua,Bernstein:1985th,Gondolo:1990dk}. However, this mechanism would not work in the case at hand for two reasons: not only were the interaction strength required to produce the DM be so large that the decay of the DM particles would proceed much too fast, but the mass required to explain the IceCube events would also be so large that the particle would be kinematically not accessible at too early times, and thus overclose the Universe.

On the other hand, in particular for very feeble interactions, freeze-in from the thermal bath is a very good alternative. In that case, the interactions of the DM particles are so weak that they never thermalise. However, they can be gradually produced at high temperatures $T\gg m_\chi$ from the primordial plasma and simply remain present in the Universe because the rate of the back-reaction is too small and the decay proceeds too slowly. In this way a sizable DM abundance can be built up, at least until the temperature in the Universe reaches the DM mass, $T \sim m_\chi$, at which point the DM particle becomes kinematically hard to access. The first reference we are aware of discussing such type of mechanism was by Langacker in 1989~\cite{Langacker:1989sv}, where freeze-in type production of sterile neutrinos has been discussed. However, the whole process was systematised and given a catchy name only much later in Ref.~\cite{Hall:2009bx}, where a frozen-in particle is called a FIMP (``Feebly Interacting Massive Particle''). Note that, unfortunately, there is an incorrect information about freeze-in present in the literature, namely that this production from the thermal plasma would actually produce a spectrum of thermal shape (i.e., Bose-Einstein or Fermi-Dirac depending on the spin of the particle), just suppressed by a momentum-independent prefactor. However, as has e.g.\ been shown in Ref.~\cite{Merle:2015vzu} for the aforementioned case of non-resonant sterile neutrino production, this is in general not true and the resulting spectrum is in fact \emph{non-thermal}. Alternatively, one can see that easily by inserting a thermal DM distribution into the equations from Ref.~\cite{Hall:2009bx}, which will clearly not be a viable solution. However, for our case the DM particle is very heavy and by that effectively act as cold DM, i.e., with non-relativistic velocities, no matter how the spectrum looks in detail.

The evolution of the number density $n_\chi$ of DM particles during the history of the Universe is described by the Boltzmann equation. It is useful to cast the Boltzmann equation in terms of the yield $Y_\chi\equiv n_\chi/\mathfrak{s}$, with $\mathfrak{s}$ being the entropy density whose expression as a function of the temperature $T$ of the thermal bath is
\begin{equation}
\mathfrak{s} = \frac{2 \pi^2}{45} g^{\mathfrak{s}}_*\left(T\right) T^3\,.
\label{eq:entropy}
\end{equation}
Here, $g^\mathfrak{s}_*\left(T\right)$ is the sum of the relativistic entropy degrees of freedom weighted by the temperatures of each species in the plasma. The Boltzmann equation reads:
\begin{equation}
\frac{dY_\chi}{dT} = -\frac{1}{\mathcal{H}\,T\,\mathfrak{s}} \left[ \frac{g_\chi}{\left(2\pi\right)^3} \int \mathcal{C} \, \frac{d^3p_\chi}{E_\chi}\right]\,,
\label{eq:Boltzmann}
\end{equation}
where the quantity in brackets contains a general collision term $\mathcal{C}$ and $\mathcal{H}$ is the Hubble parameter defined as
\begin{equation}
\mathcal{H} = 1.66 \sqrt{g_*\left(T\right)}\frac{T^2}{M_{\rm Planck}}\,,
\label{eq:hubble}
\end{equation}
where $M_{\rm Planck}$ is the Planck mass and $g_*\left(T\right)$ is the sum of the relativistic energy degrees of freedom as a function of the temperature $T$. Eq.~\eqref{eq:Boltzmann} has been obtained by assuming that the relativistic degrees of freedom of the thermal bath do not change with decreasing of the temperature, i.e.,
\begin{equation}
\frac{dg_*}{dT} = \frac{dg_*^{\mathfrak{s}}}{dT}=0\,.
\end{equation}
In our framework, this is a very good approximation, since we are interested in DM masses larger than the electroweak scale. In this regime, we can simply assume $g_*=g_*^{\mathfrak{s}}=106.75$, the value corresponding to the total number of relativistic degrees of freedom in the SM at high temperature. 

The Boltzmann equation~\eqref{eq:Boltzmann} describes how the yield $Y_\chi$ changes as a function of the temperature $T$. By integrating this equation over the temperature, or over the auxiliary variable $x\equiv m_\chi/T$, one obtains the DM relic abundance
\begin{equation}
\Omega_{\rm DM} h^2 = \frac{2 m_\chi \mathfrak{s}_0}{\rho_{\rm crit}/h^2} \left[ m_\chi \int^\infty_0 dx \frac{1}{x^2}\left(-\left.\frac{dY_\chi}{dT}\right|_{T=\frac{m_\chi}{x}}\right)\right] \,,
\label{eq:relic}
\end{equation}
where $\mathfrak{s}_0=2891.2\,{\text{cm}}^{-3}$ is today's entropy density and $\rho_{\rm crit}/h^2 = 1.054\times10^{-5}\,\text{GeV}\,\text{cm}^{-3}$ is the critical density~\cite{Agashe:2014kda}. Note that the collision term (as we will show) basically switches off for $T\gg M_S$ and $T\ll M_S$, which justifies the integration limits $0$ and $\infty$, respectively. In the above expression, the factor 2 accounts for the contribution of DM anti-particles to the relic abundance in case of Dirac DM. The result of Eq.~\eqref{eq:relic} has to be compared to the observed value of the DM relic abundance, whose $1\sigma$ range obtained by Planck~\cite{Ade:2015xua} is equal to
\begin{equation}
\left.\Omega_{\rm DM} h^2 \right|_{\rm obs} = 0.1188 \pm 0.0010\,.
\label{eq:exp}
\end{equation}
In the following we will show the processes that are involved in the DM production and report the Boltzmann equation for the settings under consideration.

\subsection{\label{sec:DM-production_diagrams}The processes behind DM production}

\subsubsection{\label{sec:DM-production_diagrams_d=4}The 4-dimensional operator}

Let us first discuss the diagrams responsible for DM production for the case of the $(d=4)$-operator of Eq.~\eqref{eq:LH}. In this scenario, the processes that provide the dominant contributions to DM production are depicted in Fig.~\ref{fig:FeynLH}. In particular, such processes are:
\begin{itemize}
\item {\it inverse decay} processes like $\nu_\alpha+H^0\rightarrow\chi$ and $\ell_\alpha+H^+\rightarrow\chi$, which occur when the DM mass $m_\chi$ is larger than $m_H+m_{\rm \nu,\ell}$ and which are weighted by $\left| y_{\alpha\chi}\right|^2$;
\item {\it Yukawa production} processes like $t+\overline{t}\rightarrow\chi+\overline{\nu_\alpha}$, whose squared matrix elements are proportional to $\left| y_{\alpha\chi}\,y_{\rm top}\right|^2$.
\end{itemize}
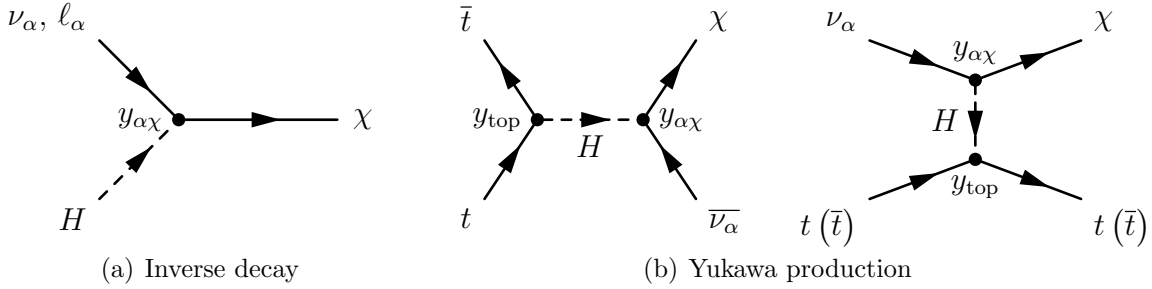
\begin{figure}[t!]
\begin{center}
\subfigure[Inverse decay]{\begin{fmffile}{lH}
\fmfframe(15,15)(15,15){
\begin{fmfgraph*}(100,60)
\fmflabel{$H$}{i1}
\fmflabel{$\nu_\alpha,\,\ell_\alpha$}{i2}
\fmflabel{$\chi$}{o1}
\fmfv{label=$y_{\alpha\chi}$}{v1}
\fmfleft{i1,i2}
\fmfright{o1}
\fmf{fermion}{i2,v1,o1}
\fmf{scalar}{i1,v1}
\fmfdotn{v}{1}
\end{fmfgraph*}}
\end{fmffile}}
\subfigure[Yukawa production]{\begin{fmffile}{ttHs}
\fmfframe(15,15)(15,15){
\begin{fmfgraph*}(100,60)
\fmflabel{$t$}{i1}
\fmflabel{$\overline{t}$}{i2}
\fmflabel{$\overline{\nu_\alpha}$}{o1}
\fmflabel{$\chi$}{o2}
\fmfv{label=$y_{\rm top}$}{v1}
\fmfv{label=$y_{\alpha\chi}$}{v2}
\fmfleft{i1,i2}
\fmfright{o1,o2}
\fmf{fermion}{i1,v1,i2}
\fmf{fermion}{o1,v2,o2}
\fmf{scalar,label=$H$}{v1,v2}
\fmfdotn{v}{2}
\end{fmfgraph*}}
\end{fmffile}
\begin{fmffile}{ttHt}
\fmfframe(15,15)(15,15){
\begin{fmfgraph*}(100,60)
\fmflabel{$t\left(\overline{t}\right)$}{i1}
\fmflabel{$\nu_\alpha$}{i2}
\fmflabel{$t\left(\overline{t}\right)$}{o1}
\fmflabel{$\chi$}{o2}
\fmfv{label=$y_{\alpha\chi}$}{v1}
\fmfv{label=$y_{\rm top}$}{v2}
\fmfleft{i1,i2}
\fmfright{o1,o2}
\fmf{fermion}{i2,v1,o2}
\fmf{fermion}{i1,v2,o1}
\fmf{scalar,label=$H$}{v1,v2}
\fmfdotn{v}{2}
\end{fmfgraph*}}
\end{fmffile}}
\end{center}
\caption{\label{fig:FeynLH}Feynman diagrams providing the dominant contribution to the DM production in case of 4-dimensional operator $\overline{L_L}H\chi$.}
\end{figure}
In general, DM particles can also be produced by other Higgs-mediated processes through all the SM Yukawa interactions. However, the dominant contribution is provided by the top quark interactions because the top Yukawa coupling $y_{\rm top}$ is $\mathcal{O}(1)$. Indeed, the other processes are negligible due to the smallness of the corresponding Yukawa couplings.

In this case, the Boltzmann equation~\eqref{eq:Boltzmann} is given by \cite{Hall:2009bx}:
\begin{equation}
\frac{dY^{(d=4)}_\chi}{dT} = \left.\frac{dY_\chi}{dT}\right|_{\rm inv.dec.} + \left.\frac{dY_\chi}{dT}\right|_{\rm Yuk.prod.} \,,
\label{eq:BoltzmannLH}
\end{equation}
where
\begin{eqnarray}
\left.\frac{dY_\chi}{dT}\right|_{\rm inv.dec.} & = & - \frac{m^2_\chi \Gamma^{(d=4)}_\chi}{\pi^2\mathcal{H}\mathfrak{s}} K_1\left(\frac{m_\chi}{T}\right) \,, \label{eq:invdec}\\
\left.\frac{dY_\chi}{dT}\right|_{\rm Yuk.prod.} & = & - \frac{1}{512\pi^6 \mathcal{H}\mathfrak{s}}\int ds\,d\Omega\, \sum_{\alpha=e,\mu,\tau}\frac{W_{t\overline{t}\rightarrow\overline{\nu_\alpha}\chi} + 2\,W_{t\nu_\alpha\rightarrow t\chi}}{\sqrt{s}}K_1\left(\frac{\sqrt{s}}{T}\right) \,.
\label{eq:Yukpro}
\end{eqnarray}
Here, $K_1$ is the first modified Bessel function of second kind. The first term on the right-hand side of the Boltzmann equation~\eqref{eq:BoltzmannLH} accounts for the {\it inverse decay} contribution, diagram (a) in Fig.~\ref{fig:FeynLH}, and it is proportional to the total decay width $\Gamma^{(d=4)}_\chi$, given by:
\begin{equation}
\Gamma^{(d=4)}_\chi = \sum_{\alpha=e,\mu,\tau}\frac{\left|y_{\alpha\chi}\right|^2}{8\pi}m_\chi\,.
\label{eq:decayLH}
\end{equation}
The second term, instead, is related to the {\it Yukawa production}, cf.\ diagrams (b) in Fig.~\ref{fig:FeynLH}. In particular, the integral of Eq.~\eqref{eq:Yukpro} is performed on the centre-of-mass energy $s$ and the solid angle $\Omega$, and the quantities $W_{ij\rightarrow kl}$ (where $i,j,k,l$ label the particles involved in the respective reaction) are related to the squared matrix elements of the corresponding processes. Their expressions are reported in App.~\ref{app:A}.

It is important to note that, in the Boltzmann equation~\eqref{eq:BoltzmannLH}, the processes destroying DM particles are negligible and need not be taken into account. This is the main characteristic of the FIMP production mechanism, where a vanishing number density (i.e., $n_\chi=0$) is generally assumed as initial condition of the Universe and where the interaction rates are suppressed by the feebleness of the SM-DM coupling. The decay rate of $\chi$ particles is also negligible, since their lifetime $\tau_\chi=\Gamma^{-1}_\chi$ has to be at least larger than the age of the Universe in order to have a DM-related signal in the IceCube detector today. 

All the previous quantities have been obtained for the case of Dirac DM particles. However, it is worth observing that, for Majorana DM particles, the final results effectively do not change, since such the factor of 2 in Eq.~\eqref{eq:relic} is absorbed by the counting of processes that contribute to the relic abundance. For instance, in the case of the inverse decay processes the factor of 2 is absorbed by the decay width $\Gamma^{(d=4)}_\chi$ that doubles for Majorana DM particles.

In general, the integral in  Eq.~\eqref{eq:relic} has to be evaluated by means of a numerical approach. However, it can be easily computed in case of the inverse decay process being the dominant contribution. In this instance, one obtains:
\begin{equation}
\left.\Omega_{\rm DM} h^2\right|_{\rm inv.dec.} = 0.1188 \left(\frac{106.75}{g_*}\right)^{3/2} \left(\frac{\sum_{\alpha=e,\mu,\tau}\left|y_{\alpha\chi}\right|^2}{7.50\times10^{-25}} \right) \,,
\label{eq:relicINVDEC}
\end{equation}
which is conveniently normalised to the observed value of DM relic abundance reported in Eq.~\eqref{eq:exp}.

\subsubsection{\label{sec:DM-production_diagrams_d=6}The 6-dimensional operator}

Let us now discuss the case of the leptophilic $(d=6)$-operator $\left(\overline{(L_{L\alpha})^C} i\sigma_2 L_{L\beta}\right) \left(\overline{\ell_{R\gamma}} \chi \right)$, which is phenomenologically equivalent to the one proposed in Ref.~\cite{Boucenna:2015tra} to explain the IceCube observations. In this case, there exist four different classes of processes (see Fig.~\ref{fig:FeynLepto} for Feynman diagrams of the last three processes are depicted explicitly, while $S$-decay would simply correspond to the ``right half'' of the leftmost diagram):
\begin{itemize}
\item \emph{decays of} $S$ \emph{particles}, $S^\pm\to\ell^\pm\chi$, which are in thermal equilibrium with the thermal bath due to the hypercharge interactions; these processes are proportional to $\left|\lambda'_\gamma\right|^2$;
\item $s$-{\it channel} processes like $\nu^c_\alpha+\overline{\ell_\beta}\rightarrow\overline{\ell_\gamma}+\chi$, whose squared matrix elements are proportional to $4\left|\lambda_{\alpha\beta}\lambda'_\gamma\right|^2$;
\item $t$-{\it channel} processes like $\nu^c_\alpha+\ell_\gamma\rightarrow\ell_\beta+\chi$ and $\overline{\ell_\beta}+\ell_\gamma\rightarrow\overline{\nu^c_\alpha}+\chi$, whose squared matrix elements are again proportional to $4\left|\lambda_{\alpha\beta}\lambda'_\gamma\right|^2$;
\item {\it annihilation} processes like $\ell_\gamma+\overline{\ell_\delta}\rightarrow\chi+\overline{\chi}$, whose squared amplitudes are weighted by $\left|\lambda'_\gamma\lambda'_{\delta}\right|^2$.
\end{itemize}
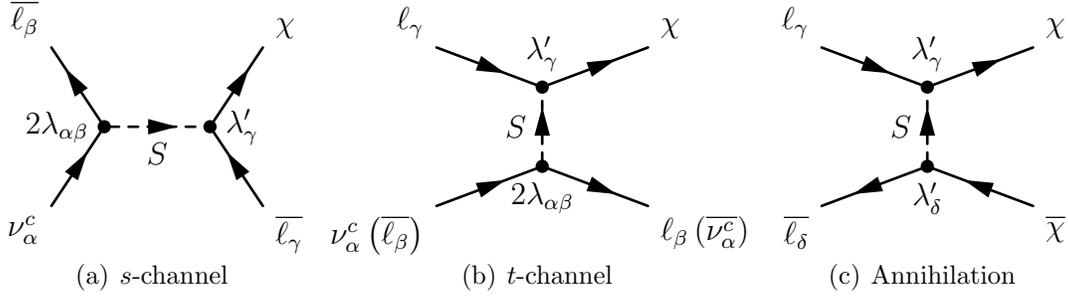
\begin{figure}[t!]
\begin{center}
\subfigure[$s$-channel]{\begin{fmffile}{Schannel}
\fmfframe(15,15)(15,15){
\begin{fmfgraph*}(100,60)
\fmflabel{$\nu^c_\alpha$}{i1}
\fmflabel{$\overline{\ell_\beta}$}{i2}
\fmflabel{$\overline{\ell_\gamma}$}{o1}
\fmflabel{$\chi$}{o2}
\fmfv{label=$2\lambda_{\alpha\beta}$}{v1}
\fmfv{label=$\lambda'_{\gamma}$}{v2}
\fmfleft{i1,i2}
\fmfright{o1,o2}
\fmf{fermion}{i1,v1,i2}
\fmf{fermion}{o1,v2,o2}
\fmf{scalar,label=$S$}{v1,v2}
\fmfdotn{v}{2}
\end{fmfgraph*}}
\end{fmffile}}
\subfigure[$t$-channel]{\begin{fmffile}{Tchannel}
\fmfframe(15,15)(15,15){
\begin{fmfgraph*}(100,60)
\fmflabel{$\nu^c_\alpha\left(\overline{\ell_\beta}\right)$}{i1}
\fmflabel{$\ell_\gamma$}{i2}
\fmflabel{$\ell_\beta\left(\overline{\nu^c_\alpha}\right)$}{o1}
\fmflabel{$\chi$}{o2}
\fmfv{label=$2\lambda_{\alpha\beta}$}{v1}
\fmfv{label=$\lambda'_{\gamma}$}{v2}
\fmfleft{i1,i2}
\fmfright{o1,o2}
\fmf{fermion}{i2,v2,o2}
\fmf{fermion}{i1,v1,o1}
\fmf{scalar,label=$S$}{v1,v2}
\fmfdotn{v}{2}
\end{fmfgraph*}}
\end{fmffile}}
\subfigure[Annihilation]{\begin{fmffile}{Ann}
\fmfframe(15,15)(15,15){
\begin{fmfgraph*}(100,60)
\fmflabel{$\overline{\ell_{\delta}}$}{i1}
\fmflabel{$\ell_\gamma$}{i2}
\fmflabel{$\overline{\chi}$}{o1}
\fmflabel{$\chi$}{o2}
\fmfv{label=$\lambda'_{\delta}$}{v1}
\fmfv{label=$\lambda'_{\gamma}$}{v2}
\fmfleft{i1,i2}
\fmfright{o1,o2}
\fmf{fermion}{i2,v2,o2}
\fmf{fermion}{o1,v1,i1}
\fmf{scalar,label=$S$}{v1,v2}
\fmfdotn{v}{2}
\end{fmfgraph*}}
\end{fmffile}}
\end{center}
\caption{\label{fig:FeynLepto}Three of the Feynman diagrams responsible for the DM production in a setting containing the 6-dimensional operator $\left(\overline{(L_{L\alpha})^C} i\sigma_2 L_{L\beta}\right) \left(\overline{\ell_{R\gamma}} \chi \right)$.}
\end{figure}
It is worth noticing that the coupling $\lambda_{\alpha\beta}$ is anti-symmetric in $\alpha$ and $\beta$, due to the structure of the operator in Eq.~\eqref{eq:sq-1}, while a factor of $2$ arises from the singlet combination of two $SU(2)$ doublets. This implies that there exist 18 different flavour combinations for the $s$-channel processes, as well as for the $t$-channel ones. On the other hand, the number of different flavour annihilation processes is 9.

Since the scalar particles $S$ carry a hypercharge equal to unity, they can interact with the SM particles through the hypercharge interactions mediated by the $U(1)_Y$ gauge boson $B_\mu$. Due to the strength of the hypercharge interactions, the $S$ particles quickly thermalise and follow a thermal distribution. Once the scalars decouple from the thermal bath and freeze-out, they can decay to SM particles or DM particles, providing a contribution to the DM relic abundance.

Indeed, according to Eqs.~\eqref{eq:sq-1},~\eqref{eq:sq-2} and~\eqref{eq:sq-3}, and the discussion in between, the scalar $S^\pm$ has at least two decay channels,
\begin{equation}
 S^\pm \to \ell_\alpha^\pm + \nu_\beta\ (\overline{\nu}_\beta) \qquad \text{and} \qquad
 S^\pm \to \ell_\gamma^\pm + \chi\ (\overline{\chi}) \quad \text{[if $M_S > \chi$]} \,,
 \label{eq:S-decay}
\end{equation}
where the second channel is of course only accessible if the mass of $S^\pm$ is larger than those of all its decay products together. However, after all it may be that $S^+$ also decays into further (e.g.\ non-SM) particles, depending on the exact model under consideration. Nevertheless, in the present paper, we focus our attention only on the most minimal setting provided by  Eqs.~\eqref{eq:sq-1},~\eqref{eq:sq-2}, and~\eqref{eq:sq-3}. In this case, the total decay width $\Gamma_S$ of $S$ particles for $M_S>m_\chi$, is explicitly given by:
\begin{equation}
\Gamma_S = \Gamma_{S\to\ell\nu} + \Gamma_{S\to\ell\chi}\,,
\label{eq:Sdecay}
\end{equation}
where
\begin{eqnarray}
\Gamma_{S\to\ell\nu} & = & \frac{1}{2\pi} \left(\sum_{\alpha=e,\mu,\tau}\sum_{\beta\neq\alpha}\left|\lambda_{\alpha\beta}\right|^2\right) M_S \,, \\
\Gamma_{S\to\ell\chi} & = & \frac{1}{8\pi} \left(\sum_{\gamma=e,\mu,\tau}\left|\lambda'_\gamma\right|^2 \right) \frac{\left(M^2_S-m^2_\chi\right)^2}{M_S\left(M^2_S+m^2_\chi\right)}\,.
\end{eqnarray}
In order to take into account the contribution of $S$ decays, we have to solve the following Boltzmann equation for $S$ particles:
\begin{equation}
\frac{dY_S}{dT} = \frac{\mathfrak{s}\left<\sigma v\right>_{\rm hyper.}}{T \, \mathcal{H}}\left[Y^2_S-\left(Y^{\rm eq}_S\right)^2\right] + \frac{\left<\Gamma\right>_{S\to\ell\nu}}{T \, \mathcal{H}} \left[Y_S - Y^{\rm eq}_S\right] + \frac{\left<\Gamma\right>_{S\to\ell\chi}}{T \, \mathcal{H}} \, Y_S \,,
\label{eq:Sboltzmann}
\end{equation}
where $Y^{\rm eq}$ is the equilibrium yield of $S$ particles. Moreover, the first term in the right-hand side of the equation is related to the hypercharge processes $S^+S^-\leftrightarrow BB$ and $S^+S^-\leftrightarrow f\overline{f}$ ($f$ stands for any SM particle), and it depends on the thermally averaged cross-section
\begin{equation}
\left<\sigma v\right>_{\rm hyper.} = \frac{2\pi \alpha^2_y}{M_S^2}\left[\frac{\,y_S^2\,\left(\sum_f n_f y_f^2 \right)}{16} +4y_S^4 \right]\left[\frac{K_1\left(M_S/T\right)}{K_2\left(M_S/T\right)}\right]^2\,,
\label{eq:sv}
\end{equation}
where $\alpha^{-1}_y = 59.008$ is the hypercharge gauge coupling at the electroweak scale,\footnote{Considering the running of the gauge coupling corresponds to a rescaling of the new couplings involved in DM production.} the quantity $y_f$ is the hypercharge of the SM multiplet $f$ ($y_S=1$), and $n_f$ is its multiplicity under the SM gauge group (e.g., $n_u=3$ for an up-quark $u$ or $n_e=1$ for an electron $e^-$). Moreover, the functions $K_1$ and $K_2$ are the first and second modified Bessel functions, respectively. The second and third terms on the right-hand side of Eq.~\eqref{eq:Sboltzmann} correspond to the processes $S^\pm\leftrightarrow \ell^\pm\nu$ and $S^\pm\to \ell^\pm\chi$, respectively.\footnote{We do not consider the inverse decay process $\ell^\pm\chi \to S^\pm$, since the number density of DM particles is negligible in the early Universe due to the feebleness of the SM-DM interactions.} In particular, we have
\begin{eqnarray}
\left<\Gamma\right>_{S\to\ell\nu} = \frac{K_1\left(M_S/T\right)}{K_2\left(M_S/T\right)} \, \Gamma_{S\to\ell\nu}& \quad\text{and}\quad & \left<\Gamma\right>_{S\to\ell\chi}  =  \frac{K_1\left(M_S/T\right)}{K_2\left(M_S/T\right)} \, \Gamma_{S\to\ell\chi} \,.
\label{eq:gamma}
\end{eqnarray}
Therefore, the Boltzmann equation for the DM particles reads
\begin{equation}
\frac{dY^{(d=6)}_\chi}{dT} = \left.\frac{dY_\chi}{dT}\right|_\text{$S$ dec.} + \left.\frac{dY_\chi}{dT}\right|_\text{$s$-ch.} + \left.\frac{dY_\chi}{dT}\right|_\text{$t$-ch.} + \left.\frac{dY_\chi}{dT}\right|_{\rm annih.}\,,
\label{eq:BoltzmannLepto}
\end{equation}
where the four contributions are related to the different processes in Fig.~\ref{fig:FeynLepto} (including $S$-decay). The first term is given by
\begin{equation}
\left.\frac{dY_\chi}{dT}\right|_\text{$S$ dec.} = - \frac{\left<\Gamma\right>_{S\to\ell\chi}}{T \, \mathcal{H}} \, Y_S \,,
\label{eq:Scontribution}
\end{equation}
while the other three terms take the form
\begin{equation}
\left.\frac{dY_\chi}{dT}\right|_i  =  - \frac{1}{512\pi^6 \mathcal{H}\mathfrak{s}}\int ds\,d\Omega\, \frac{W_i}{\sqrt{s}}K_1\left(\frac{\sqrt{s}}{T}\right) \,,
\end{equation}
where for each process the quantities $W_i$ are equal to
\begin{eqnarray}
W_\text{$s$-ch.} & = & \sum_{\alpha,\gamma=e,\mu,\tau}\sum_{\beta\neq\alpha} W_{\rm \nu^c_\alpha\overline{\ell_\beta}\rightarrow\overline{\ell_\gamma}\chi} \,, \\
W_\text{$t$-ch.} & = & \sum_{\alpha,\gamma=e,\mu,\tau}\sum_{\beta\neq\alpha} \left[ W_{\rm \nu^c_\alpha\ell_\gamma\rightarrow\ell_\beta\chi} + W_{\rm \overline{\ell_\beta}\ell_\gamma\rightarrow\overline{\nu^c_\alpha}\chi}\right] \,, \\
W_{\rm annih.} & = & \sum_{\gamma,\delta=e,\mu,\tau} W_{\rm \ell_\gamma\overline{\ell_\delta}\rightarrow\chi\overline{\chi}}\,.
\end{eqnarray}
The DM relic abundance is then obtained by plugging Eq.~\eqref{eq:BoltzmannLepto} into Eq.~\eqref{eq:relic} and numerically performing the integral over $x$. As will be shown later, the decays of $S$ particles by far provide the main contribution to the DM relic abundance in the region where the mass of scalar $S$ is larger than the DM mass, $M_S > m_\chi$. Depending on the strength of the quantities reported in Eqs.~\eqref{eq:sv}~and~\eqref{eq:gamma}, $S$ particles can freeze-out from the thermal bath or freeze-in at a temperature $T=T^*$. Therefore, if the decays of scalar mediators become efficient $\left(\left<\Gamma\right>_S > \mathcal{H}\right)$ for $T\gg T^*$, by taking $Y^{\rm eq}_S \sim0$ in Eq.~\eqref{eq:Sboltzmann} and using Eq.~\eqref{eq:Scontribution} we obtain the following analytically approximated expression for the DM relic abundance:
\begin{equation}
\left.\Omega_{\rm DM} h^2\right|_\text{$S$ dec.} \simeq \frac{2 m_\chi \mathfrak{s}_0}{\rho_{\rm crit}/h^2} \frac{\Gamma_{S\to\ell\chi}}{\Gamma_S} \, Y_S\left(T^*\right) \,.
\label{eq:relic_S0}
\end{equation}
The $s$-channel processes, instead, provide a subdominant contribution, while the contribution of the other two processes is negligible. In case of $s$-channel processes, by using the narrow width approximation according to the resonance at $T\approx M_S$, one gets the following analytical expression of the integral in Eq.~\eqref{eq:relic}:
\begin{equation}
\frac{\left.\Omega_{\rm DM} h^2\right|_\text{$s$-ch.}}{0.1188} = 
\left\{\begin{array}{lr}
\left(\frac{106.75}{g_*}\right)^{3/2} \left(\frac{\sum_{\alpha,\gamma=e,\mu,\tau}\sum_{\beta\neq\alpha}\left|\lambda_{\alpha\beta}\lambda'_\gamma\right|^2}{1.10\times10^{-21}}\right) &\quad \text{for }M_S < m_\chi \,,\\
&\\
\left(\frac{106.75}{g_*}\right)^{3/2} \left(\frac{\sum_{\alpha,\gamma=e,\mu,\tau}\sum_{\beta\neq\alpha}\left|\lambda_{\alpha\beta}\lambda'_\gamma\right|^2}{3.72\times10^{-23}}\right)\frac{\left(M^2_S-m^2_\chi\right)^2}{M_S^4}\frac{m_\chi}{\Gamma_S}& \text{for }M_S>m_\chi \,,
\end{array}\right.
\label{eq:relic_S1}
\end{equation}
where $\Gamma_S$ is the total decay width of $S$ particles and it is provided in Eq.~\eqref{eq:Sdecay}.
\begin{figure}[t!]
\begin{center}
\subfigure[$M_S<m_\chi$]{\begin{fmffile}{LeptoDecay1}
\fmfframe(15,15)(15,15){
\begin{fmfgraph*}(100,60)
\fmflabel{$\chi$}{i1}
\fmflabel{$\ell_\gamma$}{o1}
\fmflabel{$S$}{o2}
\fmfv{label=$\lambda'_{\gamma}$}{v1}
\fmfleft{i1}
\fmfright{o1,o2}
\fmf{fermion}{i1,v1,o1}
\fmf{scalar}{v1,o2}
\fmfdotn{v}{1}
\end{fmfgraph*}}
\end{fmffile}}
\subfigure[$M_S>m_\chi$ (off-shell for $M_S<m_\chi$)]{\begin{fmffile}{LeptoDecay2}
\fmfframe(15,15)(15,15){
\begin{fmfgraph*}(100,60)
\fmflabel{$\chi$}{i1}
\fmflabel{$\ell_\gamma$}{o1}
\fmflabel{$\nu_\alpha$}{o2}
\fmflabel{$\overline{\ell_\beta}$}{o3}
\fmfv{label=$\lambda'_{\gamma}$,label.angle=-100}{v1}
\fmfv{label=$\lambda_{\alpha\beta}$,label.angle=22}{v2}
\fmfleft{i1}
\fmfright{o1,o2,o3}
\fmf{fermion}{i1,v1}
\fmf{fermion}{v1,o1}
\fmf{fermion}{o3,v2,o2}
\fmf{scalar,label=$S$}{v1,v2}
\fmfdotn{v}{2}
\end{fmfgraph*}}
\end{fmffile}}
\end{center}
\caption{\label{fig:FeynLeptoDecay}DM decay channels due to the coupling leading to the effective $(d=6)$-operator.}
\end{figure}
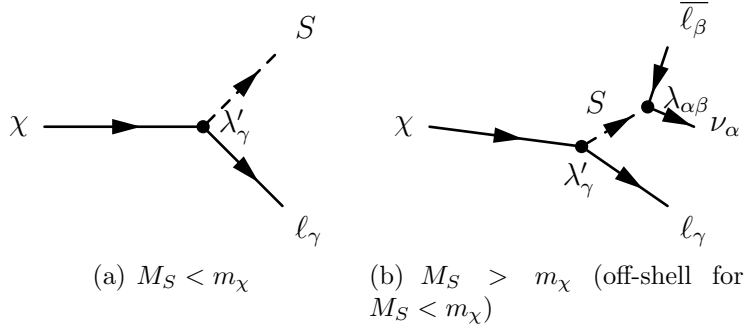
Another quantity required in this analysis is the total decay width of DM particles, due to the Zee-Babu-inspired coupling introduced above. In this framework, the $\chi$ decay processes are depicted in Fig.~\ref{fig:FeynLeptoDecay} for both cases, $M_S<m_\chi$ and $M_S>m_\chi$. The total decay width of $\chi$ particles is therefore given by
\begin{equation}
\Gamma^{(d=6)}_\chi = \left\{\begin{array}{lr}
\sum_{\gamma=e,\mu,\tau}\frac{\left|\lambda'_\gamma\right|^2}{16 \pi}m_\chi  &\quad \text{for }M_S\ll m_\chi \,,\\
&\\
\sum_{\alpha,\gamma=e,\mu,\tau}\sum_{\beta\neq\alpha}\frac{\left|\lambda_{\alpha\beta}\lambda'_\gamma\right|^2}{1536 \pi^3}\frac{m_\chi^5}{M_S^4} & \text{for }M_S \gg m_\chi \,.
\end{array}\right.
\label{eq:Leptodecay}
\end{equation}
As before, the previous expressions have been evaluated in case of Dirac DM particles, but the final results does not change in case of Majorana DM.

In the present analysis, for the sake of simplicity, we assume that the couplings are of the same order of magnitude, independently of the flavour structure. This means that
\begin{eqnarray}
\lambda_{\alpha\beta} \equiv\lambda & \quad\text{and}\quad & \lambda'_\gamma\equiv\lambda'\,.
\label{eq:eff_coup}
\end{eqnarray}
It is worth observing that, if this relation is only approximately fulfilled, each process with a different flavour structure would provide a different contribution to the DM relic abundance. In particular, in the case of large hierarchies among the couplings $\lambda_{\alpha\beta}$, only the processes proportional to larger couplings would be significant for DM production, implying that one has to take into account a smaller number of processes. On the other hand, due to neutrino oscillations, the IceCube observations are not very sensitive to different flavour structures occurring in DM decay, except for the case where $\chi\rightarrow e^+ e^- \nu_e$ is the only allowed decay channel, cf.\ Fig.~3 in Ref.~\cite{Boucenna:2015tra}.

Under the rather reasonable assumption given by Eq.~\eqref{eq:eff_coup}, as adopted in our numerical analysis, we will show in the next section that the observed DM relic abundance is obtained if $\lambda \gg \lambda'$. By using this relation, we get from Eq.~\eqref{eq:relic_S0} the following expression for the contribution of $S$ decays:
\begin{equation}
\left.\Omega_{\rm DM} h^2\right|_\text{$S$ dec.}= 0.1188 \left(\frac{\left|\lambda'\right|/\left|\lambda\right|}{4.2\times10^{-8}}\right)^2 \left(\frac{m_\chi}{\text{1 PeV}}\right)\frac{\left(M_S^2-m_\chi^2\right)^2}{M_S^4}Y_S\left(T^*\right)\,.
\label{eq:relic_S20}
\end{equation}
This quantity depends on the ratio between the two couplings $\lambda$ and $\lambda'$, since the contribution of Eq.~\eqref{eq:relic_S0} is indeed proportional to the branching ratio $\Gamma_{S\to\ell\chi}/\Gamma_S$. On the other hand, the $s$-channel contribution is solely proportional to the coupling $\lambda'$ and, for $M_S >m_\chi$, it is given by
\begin{equation}
\left.\Omega_{\rm DM} h^2\right|_\text{$s$-ch.}= 0.1188 \left(\frac{106.75}{g_*}\right)^{3/2} \left(\frac{\left|\lambda'\right|}{1.0\times10^{-12}}\right)^2 \frac{m_\chi}{M_S}\frac{\left(M_S^2-m_\chi^2\right)^2}{M_S^4}\,.
\label{eq:relic_S2}
\end{equation}

\section{\label{sec:Results}Numerical results and comparison to IceCube}

In the present analysis, the parameter spaces of both models have to be constrained by comparing the DM relic abundance to its observed value, Eq.~\eqref{eq:exp}. Moreover, two constraints on the DM lifetime $\tau_\chi$ (i.e., the inverse of total decay width) have to be taken into account: i) $\tau_\chi$ has to be larger than the age of the Universe~\cite{Audren:2014bca} ($t_{\rm Universe} \simeq 4.35\times10^{17}$~sec) and ii) it has to be compatible with the IceCube observations. In particular, the IceCube constraints on decaying DM are model-dependent, since the neutrino spectrum depends on the DM decay channels (see Refs.~\cite{Esmaili:2014rma,Aisati:2015vma}). However, the IceCube spectrum sets a lower bound on the DM lifetime of the order of $10^{28}$~sec, which is approximately model-independent.

\subsection{Results for the 4-dimensional operator}

\begin{figure}[t]
\begin{center}
\includegraphics[width=0.60\textwidth]{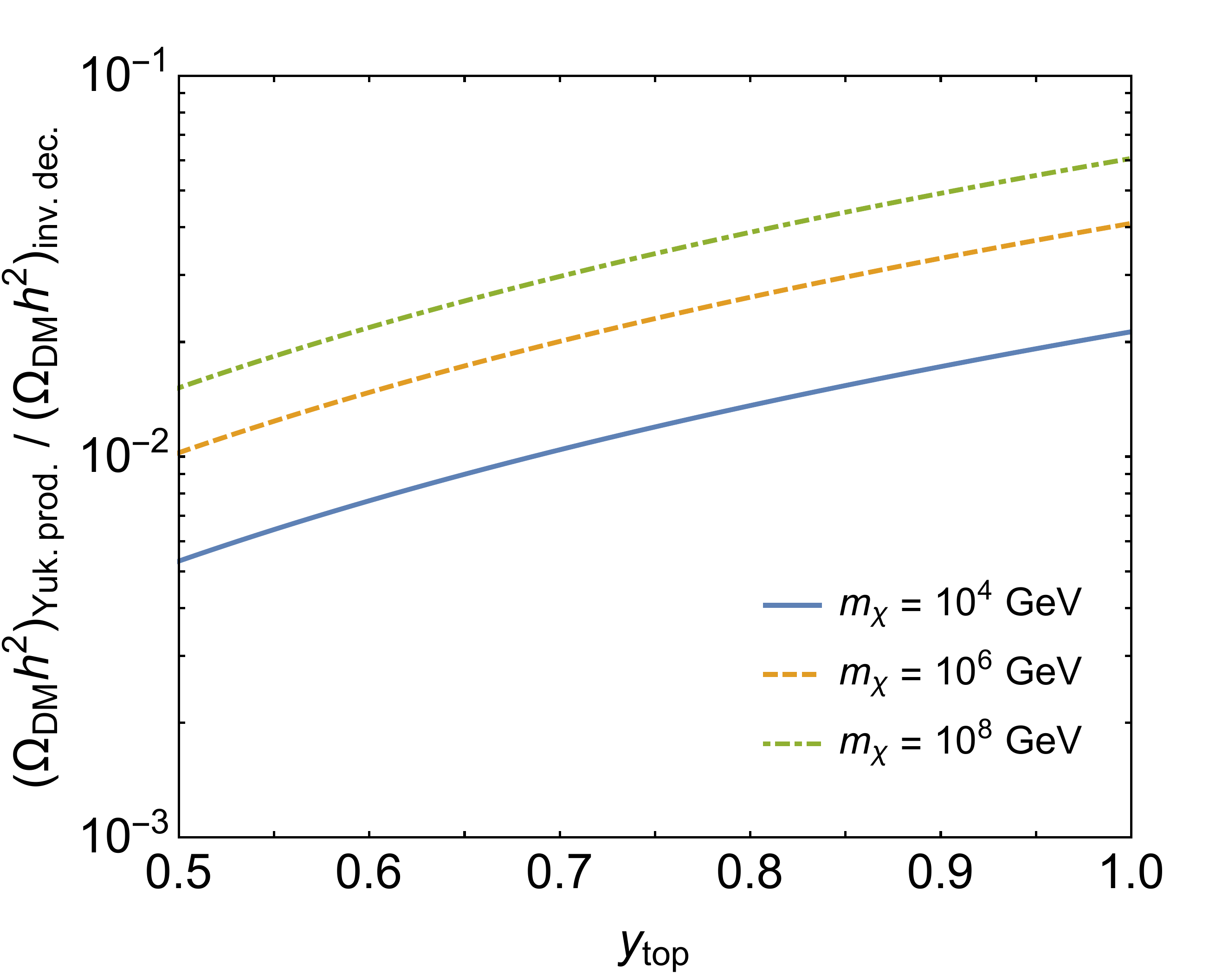}
\caption{\label{fig:1}Contributions to the DM relic abundance for the $(d=4)$-operator. The ratio between the two processes ({\it inverse decay} and {\it Yukawa production}) responsible for DM production is displayed.}
\end{center}
\end{figure}

The freeze-in DM production through the $(d=4)$-operator only, which provides a neutrino signal in IceCube at the same time, is \emph{already ruled out} by the requirement of $\tau_\chi > t_{\rm Universe}$~\cite{Audren:2014bca}. To see this, it is worth noticing that the inverse decay processes provide the dominant contribution to the DM relic abundance, as shown in Fig.~\ref{fig:1}.

In this plot we report the ratio between the Yukawa production contribution and the inverse decay one as a function of the top quark Yukawa coupling $y_{\rm top}$, for three different DM masses. Here, the coupling $y_{\rm top}$ has been considered as a free parameter since its value runs with the energy and the running depends on the high energy physics. The range considered, namely $\left[0.5,1.0\right]$, covers all the possible values obtained by SM renormalisation group equations~\cite{Degrassi:2012ry}. Therefore, Fig.~\ref{fig:1} shows that $\Omega_{\rm DM}h^2 \simeq \left.\Omega_{\rm DM}h^2\right|_{\rm inv.dec.}$ for a large region of the parameter space, implying that the DM relic abundance only depends on the couplings $y_{\alpha\chi}$ as reported in Eq.~\eqref{eq:relicINVDEC}. For any value of DM mass, the correct relic abundance is therefore obtained for
\begin{equation}
\sum_{\alpha=e,\mu,\tau}\left|y_{\alpha\chi}\right|^2 \simeq 7.50\times10^{-25}\, \quad\text{(correct relic abundance)}.
\end{equation}
However, plugging this coupling into Eq.~\eqref{eq:decayLH} implies that $\tau^{(d=4)}_\chi = \left(\Gamma^{(d=4)}_\chi\right)^{-1} \ll t_{\rm Universe}$ for $m_\chi = \mathcal{O}(1)$~PeV. Moreover, we would like to stress that, for $m_\chi\sim1$~ PeV, the squared coupling has to be
\begin{equation}
\sum_{\alpha=e,\mu,\tau}\left|y_{\alpha\chi}\right|^2 \sim\, 10^{-58}\ \ \ \text{(correct IceCube phenomenology)},
\end{equation}
in order to be compatible with the IceCube data~\cite{Higaki:2014dwa}. Thus, indeed, DM production and the IceCube high energy events cannot be brought into agreement if only the 4-dimensional operator $\overline{L_L} H \chi$ is at work.

\subsection{Results for the 6-dimensional operator}

\begin{figure}[t]
\begin{center}
\includegraphics[width=0.65\textwidth]{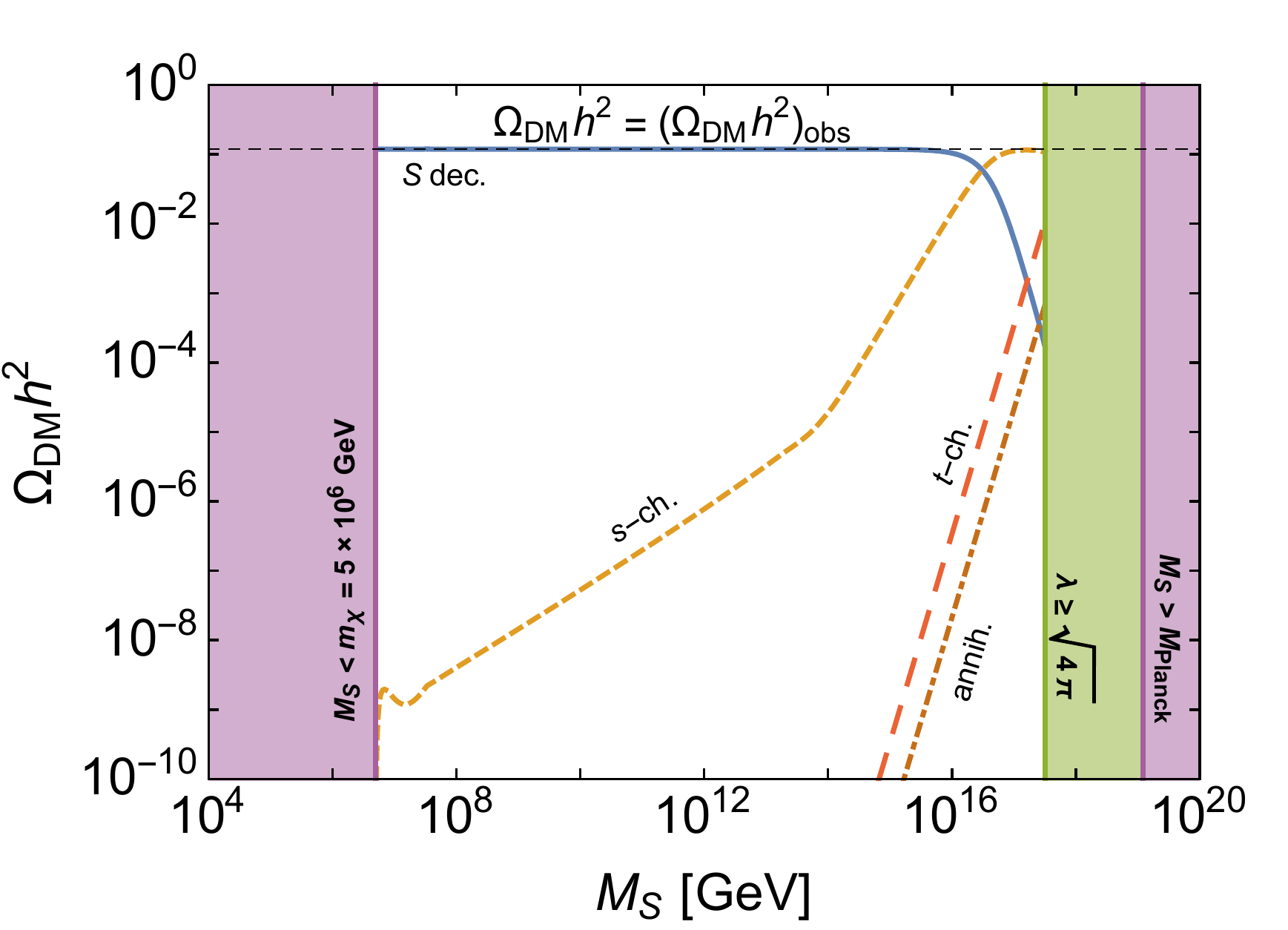}
\caption{\label{fig:1a}Contributions to the DM relic abundance for the $(d=6)$-operator as a function of the mass $M_S$. The contributions of the different processes involved in DM production are shown for $m_\chi = 5$~PeV. For each value of $M_S$, the couplings $\lambda$ and $\lambda'$ satisfy the conditions $\Omega_{\rm DM} h^2 = \left.\Omega_{\rm DM} h^2\right|_{\rm obs}$ and $\tau_\chi = 10^{28}$~sec.}
\end{center}
\end{figure}

Let us now discuss the analysis in the case of the 6-dimensional operator. All the results of this section are obtained by fixing the DM mass $m_\chi$ to be 5~PeV, i.e.\ the value of DM mass proposed to explain the IceCube PeV data in Ref.~\cite{Boucenna:2015tra}. Moreover we assume that the reheating temperature is above the mediator mass $M_S$. Fig.~\ref{fig:1a} shows the contributions of the different processes involved in the DM production as a function of the charged scalar mass, $M_S$. For each value of $M_S$, the couplings are chosen in such a way that the DM lifetime is $10^{28}$~sec and the sum of all the contributions (lines) corresponds to $\left.\Omega_{\rm DM} h^2\right|_{\rm obs}$. In the plot, the purple region on the left $\left(M_S \leq m_\chi \right)$ is not allowed due to the requirement of $\tau_\chi  = \left(\Gamma^{(d=6)}_\chi\right)^{-1} = 10^{28}$~sec. Moreover, the purple region on the right $\left(M_S > M_{\rm Planck} \right)$ displays the bound related to the Planck mass, which simply arises from our whole treatment only being valid for sub-Planckian scales. In the green region, instead, our perturbative treatment is unlikely to yield reliable results, because the observed DM abundance would require a non-perturbative coupling $\lambda$ (i.e., larger than $\sqrt{4\pi}$). 

When all constraints are satisfied, we observe that there exist two different regimes:
\begin{itemize}
\item for $M_S\lesssim 10^{15}\text{~GeV}$, the decays of thermal scalar particles provide the main contribution;
\item for $M_S \gtrsim 10^{15}\text{~GeV}$, there is a small region in which the $s$-channel processes dominate.
\end{itemize}
This means that, for small values of $M_S$, the DM relic abundance is approximately given by Eq.~\eqref{eq:relic_S20}, while for very large scalar masses it is provided by Eq.~\eqref{eq:relic_S2}. On the other hand, the other two contributions ($t$-channel and annihilation processes) are always negligible.

\begin{figure}[t]
\begin{center}
\includegraphics[width=0.49\textwidth]{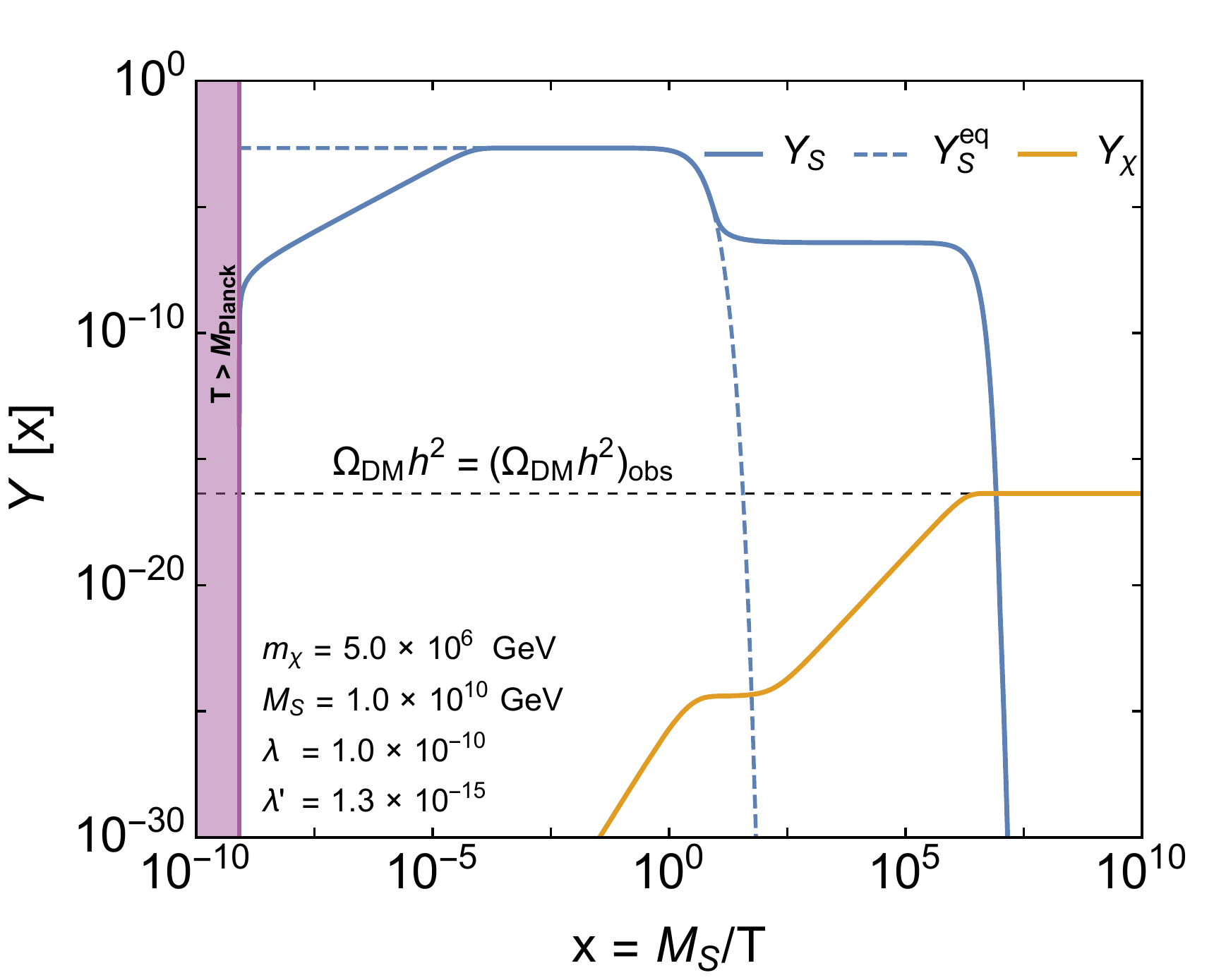}
\includegraphics[width=0.49\textwidth]{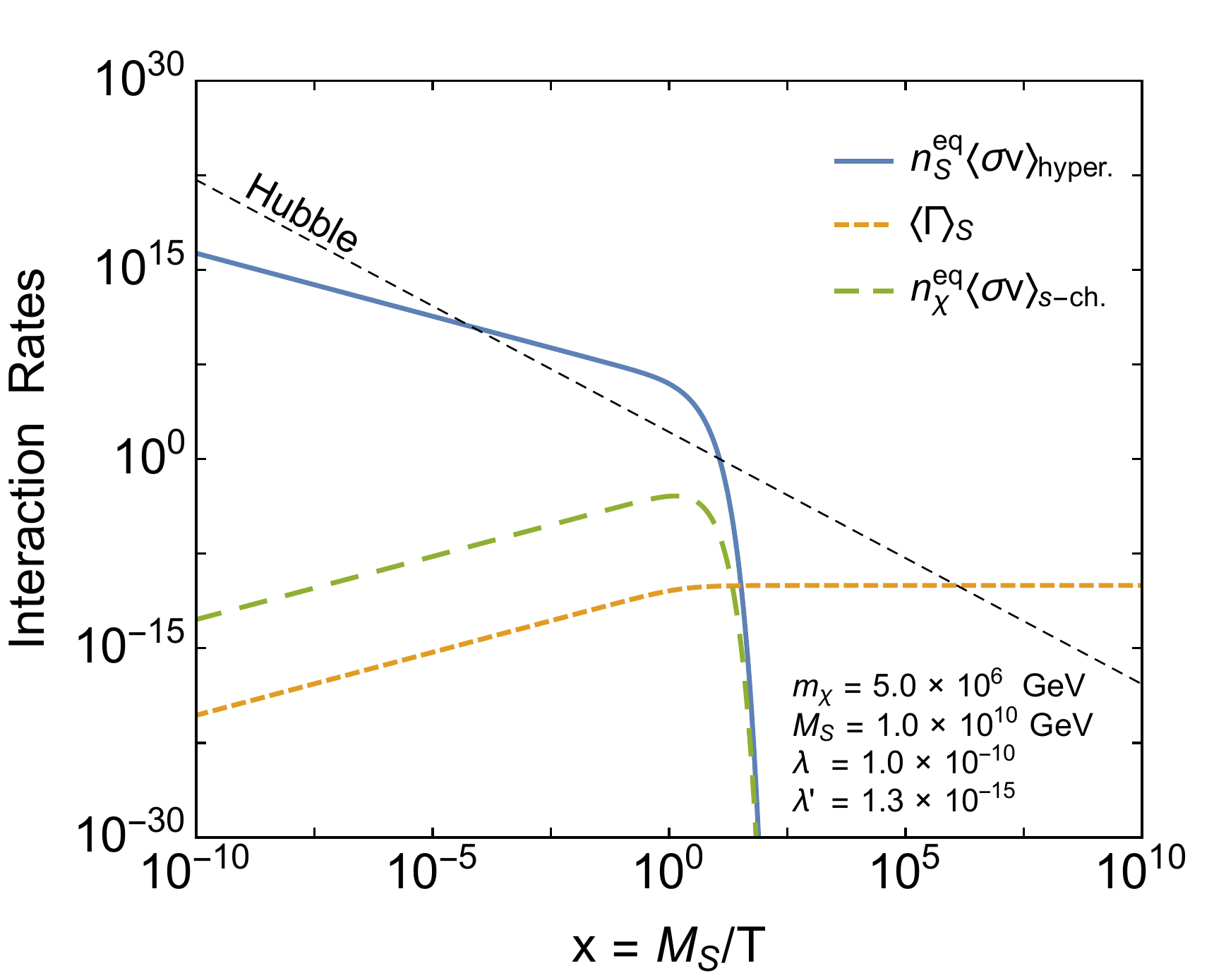}
\caption{\label{fig:2}In the left panel, we present the yields of the $S$ scalars and DM particles as a function of the auxiliary variable $x = M_S/T$. The right panel shows the interaction rates of different processes involved in the Boltzmann equations~\eqref{eq:Sboltzmann}~and~\eqref{eq:BoltzmannLepto}. In both panels, the showed quantities are evaluated for $m_\chi=5.0\times 10^6$~GeV, $M_S=1.0\times 10^{10}$~GeV, $\lambda = 1.0\times 10^{-10}$, and $\lambda' = 1.3\times 10^{-15}$.}
\end{center}
\end{figure}

In the left panel of Fig.~\ref{fig:2}, the yields $Y_S$ and $Y_\chi$ are reported as a function of the variable $x=M_S/T$ for $m_\chi =5.0$~PeV. The effective couplings $\lambda$ and $\lambda'$ defined in Eq.~\eqref{eq:eff_coup} have been fixed to $1.0\times10^{-10}$ and $1.3\times10^{-15}$, respectively, in order to obtain the correct DM relic abundance for $M_S = 10^{10}$~GeV. As one can see, for any initial distribution, the $S$ particles quickly thermalise with the thermal bath, implying that the yield $Y_S$ follows the equilibrium distribution $Y^{\rm eq}_S$ (dashed blue line in the plot). As the temperature decreases and the quantity $x$ reaches approximately the value $10$, the $S$ scalars freeze-out from the thermal bath. Then, at very low temperatures $T\sim T_{\rm dec.}$, they decay into SM and DM particles as soon as the decay rate becomes efficient $\left(\left<\Gamma\right>_S > \mathcal{H}\right)$. On the other hand, the DM yield $Y_\chi$ increases as the temperature of the bath decreases until it freezes in at $T=T_{\rm dec.}$.

The right panel of Fig.~\ref{fig:2} shows the interaction rates of different processes, i.e., the quantities $n^{\rm eq} \left<\sigma v\right>$ and $\left<\Gamma\right>$ as functions of the auxiliary variable $x$, for the same choices of the masses and couplings involved. When an interaction rate is larger than the Hubble parameter $\mathcal{H}$ (dashed black line in the plot), it means that the corresponding processes are efficient. As shown in the plot, the hypercharge interactions (solid blue line) are able to couple the $S$ particles with the thermal bath. Indeed, by comparing the two plots in Fig.~\ref{fig:2}, one can observe that the region where $n^{\rm eq}_S \left<\sigma v\right>_{\rm hyper.} \geq \mathcal{H}$ corresponds to the one where $Y_S = Y^{\rm eq}_S$, and that the scalars decouple from the thermal bath when the interaction rate equals the Hubble parameter. Moreover, the plot in the right panel displays also that the decays of $S$ particles occur once $\left<\Gamma\right>_S \approx \mathcal{H}$. On the other hand, the interaction rate of $s$-channel processes $n_\chi^{\rm eq} \left<\sigma v\right>_{s\rm{-ch.}}$, involved in the Boltzmann equation~\eqref{eq:BoltzmannLepto} of DM particles, is never larger than the Hubble parameter $\mathcal{H}$, implying that the DM particles indeed never reach the thermal equilibrium with the thermal bath -- as to be expected for FIMPs. Moreover, it firstly increases as $T$ decreases for $T>M_S$ and then rapidly falls off, because of the resonance not being met anymore for $T< M_S$. The $s$-channel contribution to the DM relic abundance corresponds to the first step in the behavior of the yield $Y_\chi$. Note that, according to Fig.~\ref{fig:1a}, the $t$-channel and annihilation processes are negligible with this choice of parameters. Furthermore, the DM particles freeze-in when $\left<\Gamma\right>_S = \mathcal{H}$ occurring at $T=T_{\rm dec.}$.

\begin{figure}[t!]
\begin{center}
\includegraphics[width=0.65\textwidth]{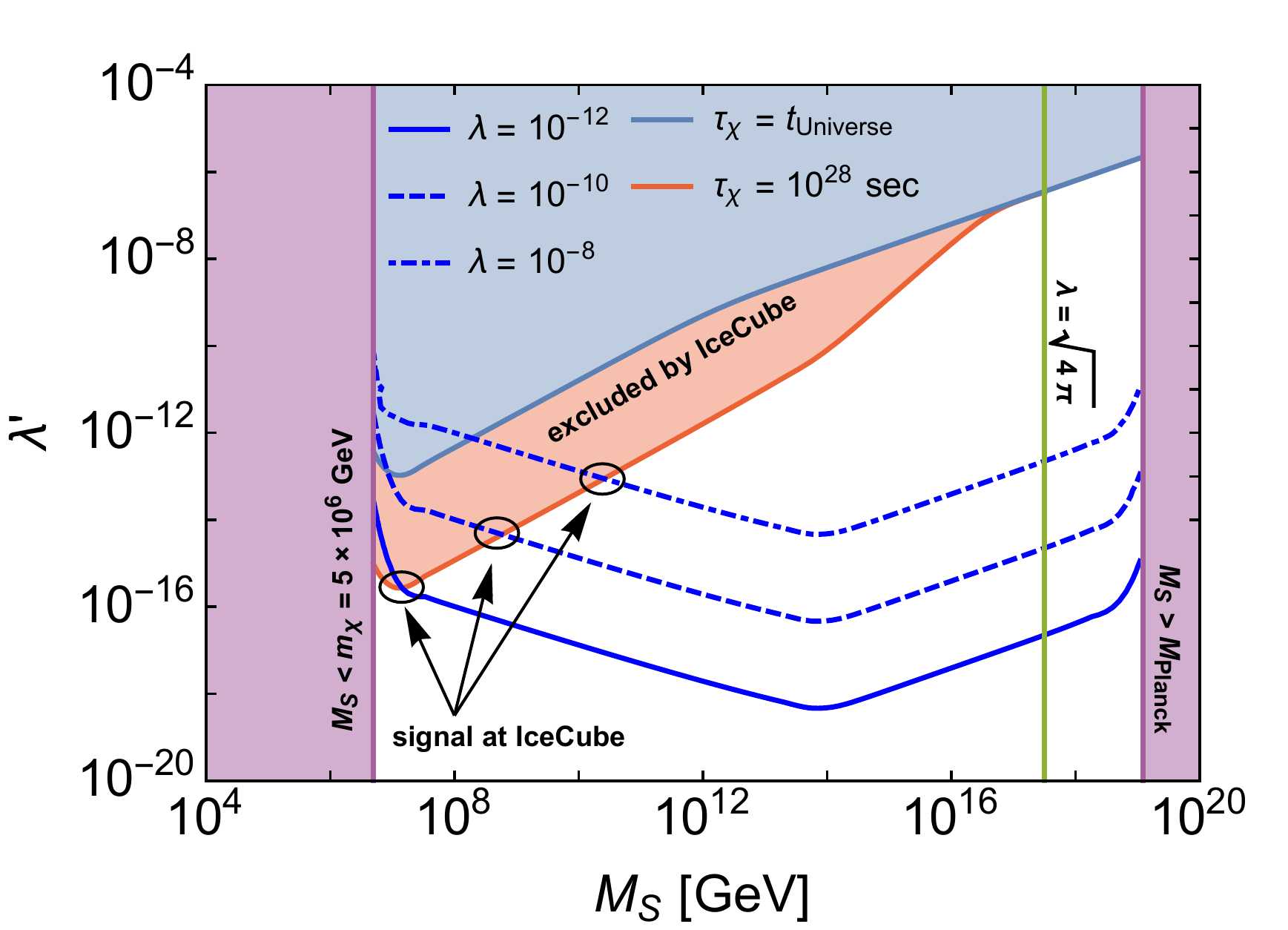}
\caption{\label{fig:3}This figure illustrates the allowed region of the parameter space for the $(d=6)$-operator, once the DM mass has been set to 5~PeV. The excluded regions are related to the constraints coming from the age of the Universe (light blue), the IceCube data (red), and the Planck and DM mass (purple). The solid red line corresponds to a DM lifetime of $10^{28}$~sec, according to Eq.~\eqref{eq:Constraint_d=6}. The solid green line bounds from below the allowed values of $M_S$ due to $\lambda$ being non-perturbative in case of $\tau_\chi = 10^{28}$~sec. The blue lines correspond to fixed values for the coupling $\lambda$. The intersections (circles) between the red line and the blue lines provide the values of $M_S$, $\lambda$ and $\lambda'$ that are compatible with the DM production and an observable signal at IceCube.}
\end{center}
\end{figure}

The main result of the present analysis is reported in Fig.~\ref{fig:3}. Here, the coupling $\lambda'$ is plotted versus the scalar mediator mass for $m_\chi=5$~PeV. The three lines shown explicitly are related to different values of the coupling $\lambda$. The light blue region is excluded since the DM lifetime would be smaller than the age of the Universe $t_{\rm Universe}$~\cite{Audren:2014bca}, whereas the red one is excluded by the requirement $\tau_\chi\gtrsim 10^{28}$~sec, according to the bounds from the neutrino spectrum observed in IceCube. This exclusion region is delimited in the plot by the solid red line, corresponding to the relation given by Eq.~\eqref{eq:Constraint_d=6}. It is worth noting that IceCube data provide the most stringent constraint on such a model of heavy DM particles. The purple regions display the bound related to the DM mass (left) and Planck mass (right), as already discussed. Thus, only the white region in the plot is allowed: it accounts for viable DM production and it is compatible with the IceCube observations and with the requirement of $M_S\leq M_{\rm Planck}$.

However, only the values of $M_S$ and $\lambda'$ surrounding to the solid red line ($\tau_\chi = 10^{28}$\,sec) are compatible with both fitting the PeV neutrinos and the DM production. The intersections (marked by the black cirlces) of the blue lines (i.e., the lines of constant $\lambda$) with the red one provide the corresponding required values for the coupling $\lambda$. As can be seen in Fig.~\ref{fig:3}, the coupling $\lambda'$ has to be smaller than $\lambda$ according to Eq.~\eqref{eq:relic_S20}. In particular, the requirement of perturbative coupling (green line in the plot) provides an {\it upper} bound on the scalar mass $M_S$ and on the couplings $\lambda'$ and $\lambda$. On the other hand, a {\it lower} bound for the values of the two couplings is in correspondence of $M_S = 1.3 \times10^7$~GeV where a minimum in $\lambda'$ is shown in the plot. In the region $m_\chi\leq M_S \leq 1.3 \times10^7\text{~GeV}$, the couplings indeed are larger than their minimum values due to the fact that the expression of Eq.~\eqref{eq:relic_S20} is proportional to the difference $\left(M_S-m_\chi\right)$. Moreover, in the case of $M_S<m_\chi$, the mass spectrum of $S$ and $\chi$ would swap, which is however not possible since it would imply a huge abundance of an electrically charged but stable particle in the Universe.

Thus, for $m_\chi=5$~PeV, the following bounds arise:
\begin{itemize}
\item Upper bound $M_S \leq 3.2\times10^{17}$~GeV and $\lambda' \leq 3.4\times10^{-7}$, according to $\lambda \leq \sqrt{4\pi}$;
\item Lower bound $\lambda' \gtrsim 2.7\times10^{-16}$ at $M_S = 1.3 \times10^7$~GeV.
\item Lower bound $\lambda \gtrsim 1.0\times10^{-13}$ for $M_S \rightarrow m_\chi$.
\end{itemize}
These bounds delimit the region of the three parameters $M_S$, $\lambda$, and $\lambda'$, whose values are compatible at the same time with the DM production and a positive IceCube signal at PeV energy.

\begin{figure}[t!]
\begin{center}
\begin{tabular}{cc}
\includegraphics[width=0.7\textwidth]{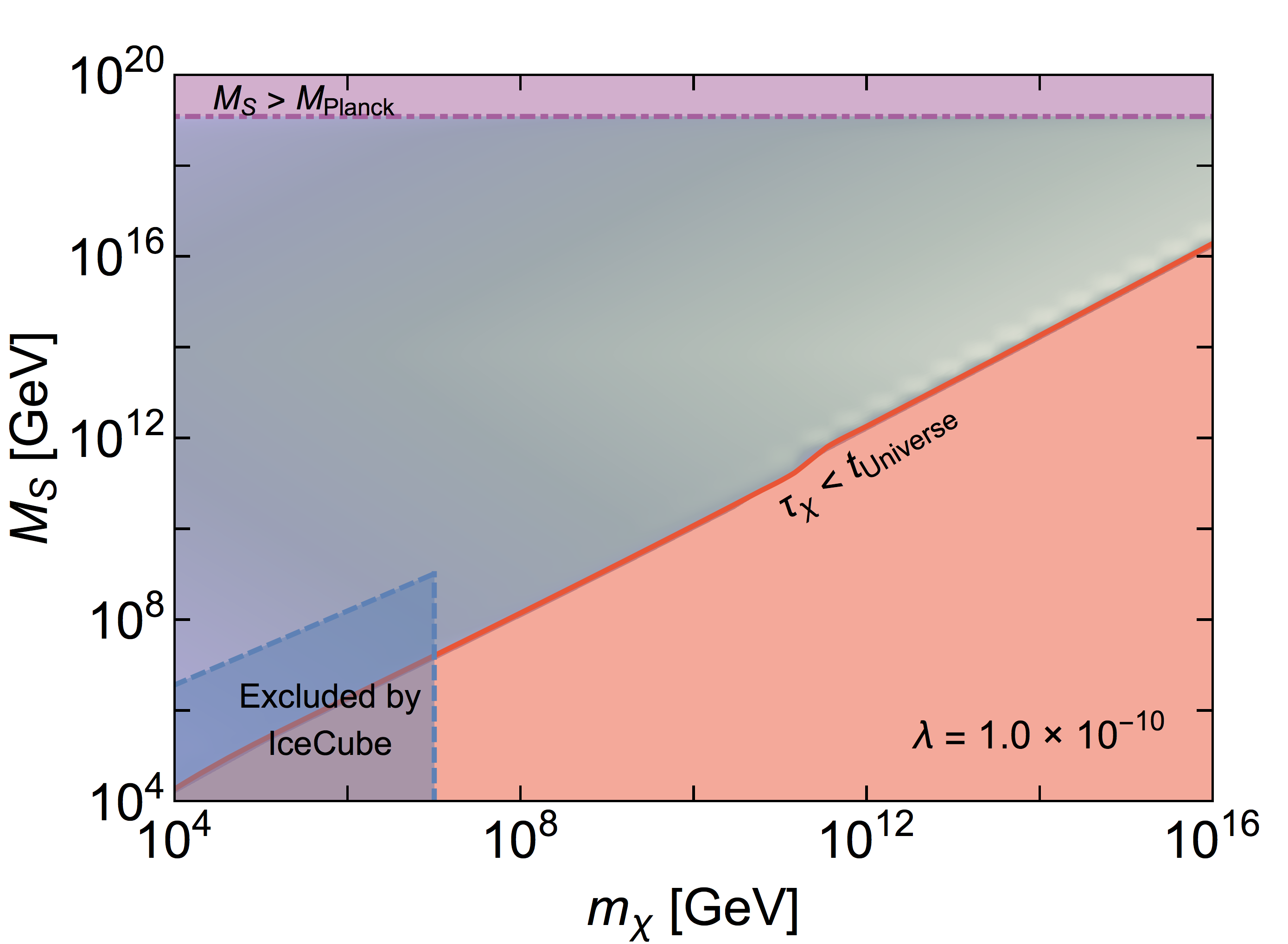} &
\begin{minipage}{2cm}
\vspace{-8.5cm}
\includegraphics[width=0.8\textwidth]{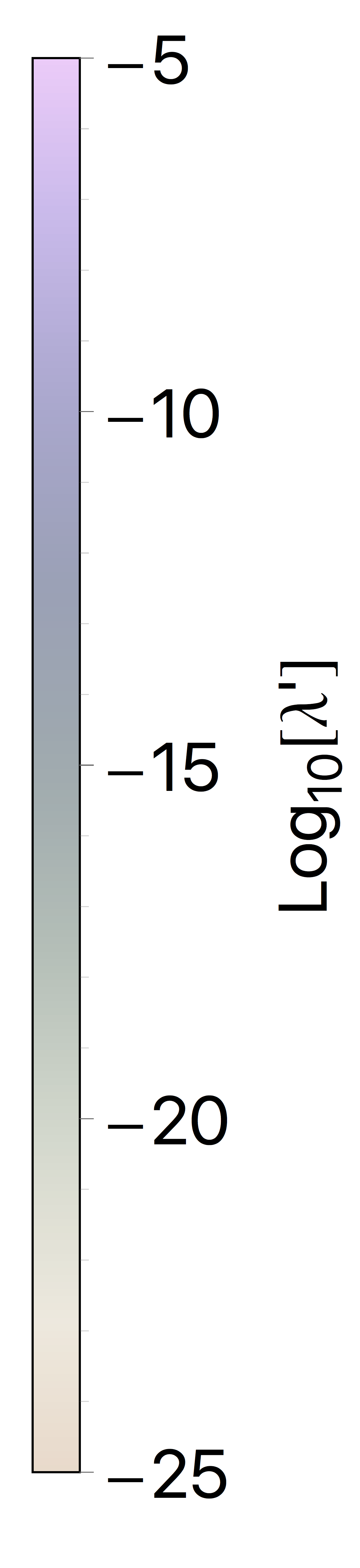}
\end{minipage}
\end{tabular}
\end{center}
\caption{\label{fig:4}Illustration of the impact of IceCube-like experiments, shown for the example of $\lambda = 10^{-10}$. The red region is excluded because of $\tau_\chi < t_{\rm Universe}$, but this constraint has not a very strong impact in practice (given that the lower right half of the plane is excluded in any case, due to $m_\chi > M_S$). As can be seen, IceCube cuts into the allowed parameter space by constraining the lifetime of the DM particle, cf. second Eq.~\eqref{eq:Leptodecay}.}
\end{figure}

Finally, in Fig.~\ref{fig:4}, we illustrate the effect of IceCube (or a similar experiment) on the parameter space. For a given value of the coupling $\lambda$ (taken to be $1.0\times 10^{-10}$ in the figure), we illustrate the $m_\chi$--$M_S$ plane with $\lambda'$ colour-coded, with the lifetime bound indicated: in the red region the DM lifetime is smaller than the age of the Universe, while in the blue region $\tau_\chi < 10^{28}$~sec, which can be considered as the lower bound for the DM lifetime coming from IceCube data. Most of this region is, however, already excluded by the requirement $m_\chi < M_S$. Glancing at the second Eq.~\eqref{eq:Leptodecay}, it is obvious that the IceCube bound can be avoided for large enough $M_S$ or small enough $\lambda'$, in both of which decrease the decay rate. It is worth observing that there exist further constraints coming from cosmological arguments like reionisation~\cite{Oldengott:2016yjc}, which are strongly model-dependent, and from other astrophysical indirect signals like gamma-rays (see for instance Refs.~\cite{Murase:2012xs,Esmaili:2015xpa}). High energy photons are indeed produced in heavy DM decays through the electroweak radiative corrections and in the interactions of charged particles with interstellar medium and the CMB.

Thus, for $\left(\overline{(L_L)^C} i\sigma_2 L_L\right) \left(\overline{\ell_R} \chi \right)$, all constraints can be met and IceCube cuts into the allowed regions, implying that future observations may probe further parameter space.

\section{\label{sec:conc}Conclusions}

In this work, we have made an attempt to answer the question whether the same operator that allows decaying Dark Matter to explain the IceCube result can also lead to successful Dark Matter production in the early Universe. We have used the two probably most generic operators describing the high energy events observed, a fundamental one with mass dimension four and an effective one with mass dimension six. We have argued that, in order to reproduce the correct lifetime needed for decaying Dark Matter, in both cases (and even more generically) some of the couplings involved in the operators need to be really small. Since the Dark Matter mass in this setting needs to be huge, this naturally seems to point towards freeze-in production in the very early Universe.

We have computed the production of Dark Matter for both operators in detail. We found that, since the 4-dimensional operator alone is not sufficient to explain both IceCube and Dark Matter production, the minimal setting has to be departed from. When using the 6-dimensional operator instead, IceCube and Dark Matter production can be explained simultaneously and the allowed parameter space left is even sufficiently small that parts of it could possibly be probed in the future.

Our results indicate that, when interpreting the IceCube high energy events as stemming from Dark Matter decay, no complicated new physics is required. Instead, a few simple additions to the Standard Model suffice to not only bring all bounds in agreement but to also provide a potentially testable parameter space left to explore. This work can be taken as motivation for future investigations of minimalistic settings, which may even be more beneficial and predictive than fully fletched models that can trivially accommodate for everything.

\section*{Acknowledgements}

We would like to thank S.~Morisi and E.~Vitagliano for helpful discussions and C.~S.~Fong for useful comments. MC acknowledges partial support by the Instituto Nazionale di Fisica Nucleare, and he thanks Georg Raffelt and the Theoretical Astroparticle Physics Group of the Max-Planck-Institut f\"ur Physik for the hospitality during the final stage of this work. AM acknowledges partial support by the European Union's Horizon 2020 research and innovation programme under the Marie Sklodowska-Curie grant agreements No.~690575 (InvisiblesPlus RISE) and No.~674896 (Elusives ITN), as well as by the Micron Technology Foundation, Inc.

\appendix
\section{\label{app:A}Appendix: Explicit expressions for $\boldsymbol{W_{ij\rightarrow kl}}$}
\renewcommand{\theequation}{A-\arabic{equation}}
\setcounter{equation}{0}  

In this appendix we report the expressions of the quantities $W_{ij\rightarrow kl}$ for all the processes that are responsible for the DM production. In particular, they take the form
\begin{equation}
W_{ij\rightarrow kl} = g_i g_j P_{ij}P_{kl} \left|\mathcal{M}\right|^2_{ij\rightarrow kl}\,,
\end{equation}
where the squared matrix element $\left|\mathcal{M}\right|^2$ is summed over initial and final spin degrees of freedom $g_i$ and averaged over initial ones, and 
\begin{equation}
P_{ij} = \frac{\left[s-(m_i+m_j)^2\right]^{1/2} \left[s-(m_i-m_j)^2\right]^{1/2}}{2\sqrt{s}} \,,
\end{equation}
where $s$ is the centre-of-mass energy and $m_i$ is the mass of particle $i$. In the following, the squared matrix elements of all the processes providing a contribution to the DM production are reported. Their expressions have been obtained by considering massless in- and out-going SM particles.

For the $(d=4)$-operator, one has to take into account the {\it Yukawa production} processes that involve the top quark, see Eq.~\eqref{eq:Yukpro}. The squared matrix elements of the two processes $t+\overline{t}\rightarrow\overline{\nu_\alpha}+\chi$ and $t+\nu_\alpha\rightarrow t+\chi$ are, respectively, given by:
\begin{eqnarray}
\left|\mathcal{M}\right|^2_{t\overline{t}\rightarrow\overline{\nu_\alpha}\chi} & = & \frac{\left|y_{\alpha\chi}y_{\rm top}\right|^2}{4} \frac{s\left(s-m_\chi^2\right)}{\left(s-M_H^2\right)^2}    \,,\\
\left|\mathcal{M}\right|^2_{t\nu_\alpha\rightarrow t\chi} & = &  \frac{\left|y_{\alpha\chi}y_{\rm top}\right|^2}{4} \frac{\left(s-m_\chi^2\right)\left[s\left(1-\cos\theta\right)+m_\chi^2\left(1+\cos\theta\right)\right]\left(1-\cos\theta\right)}{\left[\left(s-m_\chi^2\right)\left(1-\cos\theta\right)+2M_H^2\right]^2}    \,,
\end{eqnarray}
where $\theta$ is the scattering angle and $M_H$ is the mass of the SM Higgs.

For the $(d=6)$-operator, the squared amplitudes of the $s$-channel processes ($\nu^c_\alpha+\overline{\ell_\beta}\rightarrow\overline{\ell_\gamma}+\chi$), of the the $t$-channel processes ($\nu^c_\alpha+\ell_\gamma\rightarrow\ell_\beta+\chi$ and $\overline{\ell_\beta}+\ell_\gamma\rightarrow\overline{\nu^c_\alpha}+\chi$), and of the annihilation process ($\ell_\gamma+\overline{\ell_\delta}\rightarrow\chi+\overline{\chi}$) take the following forms:
\begin{eqnarray}
\left|\mathcal{M}\right|^2_{\nu^c_\alpha\overline{\ell_\beta}\rightarrow\overline{\ell_\gamma}\chi} & = & \frac{\left|\lambda_{\alpha\beta}\lambda'_{\gamma}\right|^2}{4} \frac{s\left(s-m_\chi^2\right)}{\left(s-M_S^2\right)^2}    \,,\\
\left|\mathcal{M}\right|^2_{\nu^c_\alpha\ell_\gamma\rightarrow\ell_\beta\chi} & = &  \frac{\left|\lambda_{\alpha\beta}\lambda'_{\gamma}\right|^2}{4} \frac{\left(s-m_\chi^2\right)\left[s\left(1-\cos\theta\right)+m_\chi^2\left(1+\cos\theta\right)\right]\left(1-\cos\theta\right)}{\left[\left(s-m_\chi^2\right)\left(1-\cos\theta\right)+2M_S^2\right]^2}    \,, \\
\left|\mathcal{M}\right|^2_{\overline{\ell_\beta}\ell_\gamma\rightarrow\overline{\nu^c_\alpha}\chi} & = &  \left|\mathcal{M}\right|^2_{\nu^c_\alpha\ell_\gamma\rightarrow\ell_\beta\chi}  \,,\\
\left|\mathcal{M}\right|^2_{\ell_\gamma\overline{\ell_\delta}\rightarrow\chi\overline{\chi}} & = & \frac{\left|\lambda'_{\gamma}\lambda'_{\delta}\right|^2}{4} \frac{s\left[ s\left(1+\cos^2\theta\right) - 4m_\chi^2 \cos^2\theta - 2 \sqrt{s\left(s-m_\chi^2\right)}\cos\theta \right]}{\left[s - \sqrt{s\left(s-m_\chi^2\right)}\cos\theta +2\left(M_S^2 - m_\chi^2\right)\right]^2} \,.
\end{eqnarray}

\bibliographystyle{JHEP}
\bibliography{FIMP-IC}

\end{document}